\documentclass[lettersize,journal]{IEEEtran}
\usepackage{amsmath,amsfonts}
\usepackage{algorithmic}
\usepackage{algorithm}
\usepackage{array}
\usepackage[caption=false,font=normalsize,labelfont=sf,textfont=sf]{subfig}
\usepackage{textcomp}
\usepackage{stfloats}
\usepackage{url}
\usepackage{makecell}
\usepackage{verbatim}
\usepackage{graphicx}
\usepackage{cite}
\usepackage{soul}
\usepackage{multirow}
\usepackage{mathtools}
\usepackage{color,soul}
\usepackage{breqn}
\begin{document}

\newcommand{\e}{\mathcal{E}}
\newcommand{\probP}{\text{I\kern-0.15em P}}

\title{DECOFFEE: Decentralized Reinforcement Learning for Time-critical Workload Offloading and Energy Efficiency across the Computing Continuum}

\author{Anastasios E. Giannopoulos,~\IEEEmembership{Member,~IEEE,} Sotirios T. Spantideas, and Panagiotis Trakadas

\thanks{Corresponding author: Anastasios Giannopoulos}%
\thanks{Anastasios Giannopoulos is with the Research \& Development Department, Four Dot Infinity, Chalandri, Athens, P.C. 15231, Greece (e-mail: angianno@fourdotinfinity.com).}%
\thanks{Sotirios Spantideas is with the Research \& Development Department, Four Dot Infinity, Chalandri, Athens, P.C. 15231, Greece (e-mail: sospanti@fourdotinfinity.com).}%
\thanks{Panagiotis Trakadas is with the Department of Ports Management and Shipping, National and Kapodistrian University of Athens, Psachna, Evia, P.C. 34400, Greece (e-mail: ptrakadas@uoa.gr).}%
\thanks{Digital Object Identifier XX.YYYY/TMC.ZZZZ.LLLLLL}}%

\markboth{PREPRINT}%
{Giannopoulos \MakeLowercase{\textit{et al.}}: DECOFFEE: Decentralized Reinforcement Learning for Time-critical Workload Offloading and Energy Efficiency across the Computing Continuum}


\maketitle

\begin{abstract}
The rapid proliferation of latency-sensitive and battery-constrained Internet-of-Things (IoT) applications has intensified the need for intelligent workload placement mechanisms across the Edge–Cloud computing continuum. In such environments, far-edge nodes must dynamically decide whether to execute workloads locally or offload them to neighboring nodes or the cloud, while accounting for execution delay, energy consumption, and strict timeout constraints. However, workload placement in large-scale distributed infrastructures is a highly dynamic and non-convex optimization problem due to stochastic arrivals, heterogeneous computing capacities, and time-varying network conditions. This paper proposes DECOFFEE, a decentralized reinforcement learning framework for time-critical workload offloading and energy-efficient operation across the computing continuum. The proposed multi-agent learning scheme jointly optimizes system delay, energy consumption, and workload drop rate through adaptive placement decisions. Each edge agent operates as an autonomous learning entity that derives an optimal policy from local system observations and predicted network conditions. The workload placement process is formulated as parallel Markov Decision Processes and solved using a Double Dueling Deep Q-Network (DQN) architecture enhanced with Long Short-Term Memory (LSTM) forecasting to anticipate future load conditions. Extensive simulations demonstrate that DECOFFEE and its variants consistently outperform conventional rule-based and heuristic placement strategies, achieving significant reductions in delay, energy consumption, and workload drop rate under varying traffic and network conditions.
\end{abstract}

\begin{IEEEkeywords}
Cloud-edge computing, Computing continuum, Deep reinforcement learning, Edge intelligence, Energy efficiency, Task offloading, Workload placement
\end{IEEEkeywords}

\section{Introduction}

The rapid expansion of Internet-of-Things (IoT) systems has brought forth a new class of time-sensitive and resource-constrained applications, including real-time video analytics, autonomous navigation, industrial monitoring, and extended reality services \cite{al2024computing}. These applications impose stringent requirements on latency, reliability, and energy efficiency, which cannot be satisfied solely through centralized cloud computing models due to issues such as high latency and bandwidth limitations. As a result, modern compute architectures are evolving toward a distributed paradigm, where edge nodes and cloud resources are jointly leveraged to form an IoT-to-Edge-to-Cloud computing (IECC) continuum \cite{giannopoulos2024placing}. Within this continuum, a critical challenge lies in how to dynamically offload and execute computational workloads generated by IoT endpoints. These workloads arrive continuously, often with tight deadlines and limited tolerance for delay or energy overhead. At the same time, edge and cloud resources exhibit varying capacity, dynamic availability, heterogeneous capabilities, and variable levels of congestion \cite{alnoman2019emerging}. Hence, workload placement decisions must account for spatiotemporal fluctuations in system status, while ensuring that critical tasks are completed within prescribed latency and energy budgets.

Compounding this complexity is the inherently decentralized nature of the continuum, where IoT devices and edge agents must often make decisions with only partial or localized knowledge of the global system state. Furthermore, reactive strategies that rely solely on current observations are prone to suboptimal behavior due to the delay between decision execution and system feedback. This gap highlights the importance of proactive and predictive workload offloading policies that anticipate future system conditions, particularly node load and energy usage, to better align workload placement with real-time constraints \cite{liu2024learning}. 

\subsection{Workload Placement across the Edge-Cloud Continuum}\label{sec:edge_cloud}

In distributed computing architectures, the Edge–Cloud Continuum (ECC) is expected to support a range of functionalities including real-time data processing, workload orchestration, energy-aware scheduling, and cooperative task sharing between nodes \cite{donta2023exploring}. Compute agents located at the edge of the network, often referred to as edge aggregators or edge agents, act as intermediaries between the data-producing IoT devices and the high-capacity cloud. These agents must operate under stringent latency and energy constraints while adapting to fluctuations in workload intensity and system resources \cite{giannopoulos2024hoodie}. Hence, a key function within this paradigm is the dynamic workload placement, defined as the decision-making process for assigning incoming computational tasks to either local processing units, neighboring edge nodes (horizontal offloading), or remote cloud servers (vertical offloading). Unlike static placement schemes, dynamic workload placement must react to real-time observations and future expectations of node load, network conditions, and application deadlines \cite{asim2020review}. The latter involves a complex trade-off between executing locally a workload or offloading it for remote processing. On the one hand, executing a task locally may reduce communication delay but, on the contrary, it can increase local congestion or energy consumption. Conversely, offloading to other nodes can incur transmission overhead or violate time constraints due to queuing and execution delays at the destination.

To make informed placement decisions, it is essential to account for the full range of delay components, including queuing latency at both workload source/destination, wireless and wired transmission time, and execution delay at the destination node \cite{wang2016dynamic}. Simultaneously, energy consumption must be modeled holistically, incorporating local processing usage, offloading transmission cost, and the energy footprint of processing workloads at external nodes. These elements interact in a tightly coupled manner, leading to a non-convex multi-objective optimization problem, which involves jointly minimizing workload latency and energy usage while ensuring successful completion within hard deadlines and reliability constraints (e.g., low number of workload drops) \cite{gao2022deadline}.

This formulation becomes even more challenging in distributed settings where decision-making must occur under partial observability, asynchronous and parallel workload arrivals, and prediction uncertainty. Thus, efficient workload placement policies within ECC must be both adaptive and anticipatory, capable of leveraging forecasting mechanisms and local feedback to continuously refine their offloading strategies \cite{nishad2025adaptive}. Recent advances in Machine Learning (ML) and Deep Reinforcement Learning (DRL) have shown promise in enabling adaptive task offloading strategies in such environments \cite{li2024computation}. DRL agents can learn optimal policies by interacting with the environment and balancing immediate and long-term trade-offs across multiple objectives. However, the distributed nature of ECC forces the deployment of large-scale distributed DRL agents so as to ensure model scalability (i.e., heavy centralized DRL agents are avoided), local energy-awareness, and delay guarantees under partial observability and parallel execution dynamics \cite{giannopoulos2024pdppnet}. Based on the above, in ECC, the workload offloading problem may be treated as a decentralized decision-making process, where each edge compute node independently learns to assign workloads across local, edge, and cloud resources under predictive telemetry across the continuum and constraints for time criticality and energy efficiency (EE).

\subsection{Related Work}\label{related_work}

Initial approaches to workload offloading in wireless and edge computing environments primarily focused on traditional methods, encompassing centralized optimization algorithms and heuristic-based schemes. These techniques often relied on extensive knowledge of the network state, resource availability, and workload characteristics to make offloading decisions. For instance, Luo et al. \cite{luo2021resource} studied offloading problems within a green and sustainable Mobile Edge Computing (MEC) framework, often employing the Lyapunov technique to minimize response time. While Lyapunov optimization enables online decision-making, it typically yields approximately optimal or sub-optimal performance \cite{luo2021resource}. Similarly, complex resource allocation problems were formulated as Mixed-Integer Programs (MIPs) and solved via decomposition methods. However, such MIPs are inherently NP-hard, meaning exact solutions incur exponential computational complexity, making them impractical for real-time decision-making in dynamic environments. Moro et al. \cite{moro2021joint} formulated resource allocation as a constrained optimization problem to jointly manage compute and radio resources in MEC, aiming to derive optimal allocations. Yet, solving these complex constrained optimization problems requires extensive computational resources and a global view of the system, hindering their applicability in rapidly changing scenarios. Furthermore, Bute et al. \cite{bute2021efficient} highlighted that task allocation in vehicular edge computing networks is an NP-hard 0/1 Knapsack problem. They proposed heuristic algorithms to achieve real-time decision-making \cite{bute2021efficient}. Wang et al. \cite{wang2021dependent} noted that many existing solutions using heuristic or approximation methods for task offloading require hand-tuned adjustments that are expensive, time-consuming, and impractical in dynamic MEC scenarios. This implies a reactive rather than proactive approach, where policies must be re-executed with significant computational overhead in each time slot as the environment changes. Kalinagac et al. \cite{kalinagac2023prioritization} similarly acknowledged the NP-hard nature of task assignment in Unmanned Aerial Vehicle (UAV)-assisted edge networks and proposed heuristic and quasi-optimal algorithms, which, by their nature, do not guarantee global optimality and can still be computationally demanding for frequent re-evaluation. These traditional methods generally struggle with the dynamic, heterogeneous, and distributed nature of the computing continuum, often exhibiting polynomial complexity, requiring constant re-execution, and lacking proactive capabilities to adapt to unforeseen changes.

To overcome the limitations of traditional approaches, recent research has increasingly explored ML and DRL techniques for dynamic task offloading in 5G edge-cloud environments and beyond. Tang et al. \cite{tang2020deep} proposed a model-free DRL-based distributed algorithm incorporating traffic prediction, dueling Deep Q-Network (DQN), and double-DQN to allow devices to make decentralized offloading decisions without needing to know other devices' task models. This approach aimed to minimize long-term cost under uncertain load dynamics and delay-sensitive tasks. However, while distributed at the execution level, its decentralized nature might not fully address global coordination challenges or partial observability of the overall system state. Nieto et al. \cite{nieto2024deep} introduced a distributed DRL tool to optimize binary task offloading decisions, focusing on maximizing Quality-of-Experience (QoE) while satisfying latency requirements. They also acknowledged that traditional ML approaches often struggle due to the lack of representative datasets for every possible environment state, which DRL overcomes by learning through continuous interaction \cite{nieto2024deep}. Dai et al. \cite{dai2020edge} utilized DRL to design an optimal computation offloading and resource allocation strategy for minimizing system energy consumption in multi-user end-edge-cloud orchestrated networks, modeling the problem as a Markov Decision Process (MDP). While effective for energy reduction, their solution implies a more centralized DRL agent managing the overall system. Yang et al. \cite{yang2024beyond} explored decentralized frameworks for MEC offloading, noting that such frameworks can lead to multi-agent architectures that scale better. They discussed multi-agent methods where agents make independent decisions, but often these works ignored the influences caused by other agents’ decisions, which could hinder convergence to a globally optimum solution. Similarly, Wang et al. \cite{wang2021dependent} also highlighted DRL ability to learn efficient offloading decisions by interacting with the environment, especially for dependent tasks represented by Directed Acyclic Graphs (DAGs). Yet, many DRL-based solutions still tend towards centralized training or single-objective optimization, which, despite offering adaptivity, may not fully capture the complexities of truly decentralized and multi-objective trade-offs under partial observability.

Despite advancements, current literature often overlooks the horizontal and vertical offloading complexities inherent in multi-MEC and ECC environments, where tasks can be distributed among neighboring MEC sites or deeper into the vehicular fog \cite{wakgra2024multi}. For example, some studies have focused on optimizing task offloading for non-divisible and delay-sensitive tasks using model-free DRL-based distributed algorithms \cite{yamansavascilar2022deepedge}, while others have explored joint optimization of device-level and edge-level task offloading to minimize execution delay and energy consumption. However, these approaches frequently rely on centralized training structures, even if execution is distributed, which can limit scalability and adaptability in highly dynamic and decentralized computing continuum environments \cite{yang2024beyond}. Also, significant gaps persist in both traditional and ML/DRL-based offloading strategies, particularly regarding the joint optimization of multiple conflicting objectives (e.g., workload execution latency, energy consumption, and workload drop rate due to deadline violations) in a coordinated yet decentralized manner. Furthermore, existing approaches often neglect proactive decision-making, leading to suboptimal or outdated actions as environments change. 

\subsection{Contributions}\label{contributions}

In this work, a decentralized RL for time-critical workload offloading and energy
efficiency across the ECC (DECOFFEE) is proposed to address the above challenges. This is achieved by formulating the workload offloading problem as parallel agent-specific MDPs, enabling fully decentralized decision-making by each edge agent within a shared stochastic environment. DECOFFEE leverages a model-free DRL approach with a Long Short-Term Memory (LSTM)-enhanced DQN to handle temporal dynamics, partial observability, and the stochastic and non-stationary nature of the continuum. This design incorporates forecasting and local feedback for proactive decision-making and can offer enhanced scalability, adaptability, and generalizability across heterogeneous continuum environments, providing a dynamic solution for time-critical and energy-efficient workload offloading.

The main contributions of this work are summarized as follows. First, we propose a joint multi-objective optimization framework (DECOFFEE) that enables decentralized RL agents to make time-critical and energy-aware offloading decisions by jointly optimizing workload execution latency, energy consumption, and workload drop rate due to deadline violations. Based on the decentralized multi-agent DRL paradigm, we formulate the workload offloading process as a collection of parallel agent-specific MDPs, allowing each edge agent to autonomously determine both horizontal and vertical offloading actions within a shared stochastic environment. Latency and energy consumption components are mathematically formulated across all stages of workload placement (queueing, processing, and transmission) within the ECC. Also, the model-free and adaptive learning employed in this work obviates the need for explicit modeling of global system dynamics or inter-agent dependencies, thus offering enhanced scalability and generalizability across diverse and heterogeneous continuum environments. In addition, LSTM-aided DQN predictions are integrated within each DECOFFEE agent, enabling a forecasting module for predictive state enrichment and a dueling DQL module for efficient policy learning. Finally, DECOFFEE supports lightweight and real-time deployability with low inference cost, making it highly suitable for online operation in real-time edge computing systems where rapid decision-making for workload placement is paramount.

The remainder of this paper is organized as follows. Section ~\ref{sec:system_model} presents the system model of the computing continuum and describes the workload generation, placement mechanisms, and delay and energy consumption models. Section~\ref{sec:task_offloading} formulates the decentralized workload offloading problem as a set of parallel MDPs and introduces the distributed DRL formulation adopted in DECOFFEE. Section~\ref{sec:decoffee} describes the proposed DECOFFEE algorithm, including the LSTM-based forecasting module and the Double Dueling Deep Q-Network architecture used for policy learning. Section~\ref{sec:experimental_results} evaluates the performance of the proposed framework through extensive simulations and comparative analysis against representative baseline workload placement strategies. Finally, Section~\ref{sec:conclusions} concludes the paper and outlines future research directions.

To enhance readability, Table \ref{table1} lists all the acronyms used in this article.

\begin{table}
\centering
\caption{Acronyms}
\label{table1}
\setlength{\tabcolsep}{3pt}
\begin{tabular}{|p{55pt}||p{160pt}|}
\hline
\textbf{Acronym} & \textbf{Meaning} \\
\hline
A\&V & Advantage \& Value layer\\
CA & Cloud Agent\\
CFS & Completely Fair Scheduling\\
CPU & Central Processing Unit\\
DAG & Directed Acyclic Graph\\
DECOFFEE & \textbf{DEC}entralized RL for time-critical workload \par \textbf{OFF}loading and \textbf{E}nergy
\textbf{E}fficiency across the \par Edge-Cloud Continuum\\
DQL & Deep Q-Learning\\
DQN & Deep Q-Network\\
DRL & Deep Reinforcement Learning\\
EA &  Edge Agent\\
ECC &  Edge-Cloud Continuum\\
EE & Energy Efficiency\\ 
FIFO & First In First Out\\
ID & Identifier\\
IECC & IoT-Edge-Cloud Continuum\\
IoT & Internet of Things\\
LSTM & Long Short-Term Memory\\
MDP & Markov Decision Process\\
MEC & Mobile Edge Computing\\
MIP & Mixed-Integer Programming\\
ML & Machine Learning\\
MSE & Mean Squared Error\\
QoE & Quality of Experience\\
ReLU & Rectified Linear Unit\\
RL & Reinforcement Learning\\
RU & Radio Unit\\
TA & Telemetry Agent\\
UAV & Unmanned Aerial Vehicle\\
WS & Workload Stack\\
\hline
\end{tabular}
\label{tab1}
\end{table}
%
%
%
%
\section{System Model}\label{sec:system_model}

This section presents a detailed overview of the system architecture, operational framework, and functional components that form the basis of the proposed decentralized computing environment. We first introduce the general system model, followed by a description of each core element and module involved. For clarity and consistency, all mathematical symbols used in the following methodology are compiled in Table~\ref{tab2}.

\begin{table*}[!t]
\caption{Mathematical Symbols}
\label{tab2}
\centering
\begin{tabular}{|c|p{3.9cm}||c|p{3.9cm}||c|p{3.9cm}|}
\hline
\textbf{Symbol} & \textbf{Description} & \textbf{Symbol} & \textbf{Description} & \textbf{Symbol} & \textbf{Description}\\

\hline
$\mathcal{N}'$ & Set of Edge-Cloud nodes 
& $\mathcal{N}$ & Set of EAs
& $\mathcal{M}$ & Set of TAs \\

\hline
$\mathcal{T}$ & Set of time slots
& $\mathcal{R}$ & Set of link data rates
& $\mathcal{H}$ & Set of workload sizes \\

\hline
$N$ & Number of EAs
& $M$ & Number of TAs
& $T$ & Number of time slots \\

\hline
$R_{n,k}^H$ & Horizontal link data rate [bps]
& $R_{n,k}^V$ & Vertical link data rate [bps] 
& $\Delta$ & Time slot duration [sec] \\

\hline
$\mathcal{P}$ & Workload arrival probability
& $w_n(t)$ & Workload assignment number 
& $z_n(t)$ & Workload arrival index \\

\hline
$\eta_n(t)$ & Size of $w_n(t)$ [bits]
& $T_n^{\text{max}}(t)$ & Timeout of $w_n(t)$ [time slot]
& $\rho_n(t)$ & Processing density of $w_n(t)$ [CPU cycles/bit] \\

\hline
$x_n(t)$ & Local compute decision for $w_n(t)$
& $y_{n,k}(t)$ & Offloading decision for $w_n(t)$ 
& $\boldsymbol{D}_n(t)$ & Placement decision for $w_n(t)$ \\

\hline
$\tau_n^{priv}(t)$ & Waiting time of $w_n(t)$ in private WS [time slots]
& $\psi_n^{priv}(t)$ & Completion time slot of $w_n(t)$ placed in the private WS 
& $f_{n}^{EA,priv}$ & Private CPU processing capacity in EA $n$ [cycles/sec] \\

\hline
$\tau_n^{off}(t)$ & Waiting time of $w_n(t)$ in offloading WS [time slots]
& $\psi_n^{off}(t)$ & Completion time slot of $w_n(t)$ placed in the offloading WS 
& $R_{n,k}$ & Data rate between EA $n$ and node $k$ [bps]\\

\hline
$w_{n,k}^{pub}(t)$ & Workload ID offloaded by EA $n$ to node $k$
& $\eta_{n,k}^{pub}(t)$ & Size of $w_{n,k}^{pub}(t)$ [bits]
& $l_{n,k}^{pub}(t)$ & Length of public WS $n$ of node $k$ [bits] \\

\hline
$\psi_{n,k}^{pub}(t)$ & Completion time slot of $w_{n,k}^{pub}(t)$
& $m_{n,k}^{pub}(t)$ & Number of bits removed by public WS $n$ of EA $k$ [bits]
& $f_n^{pub}$ & Public CPU processing capacity of node $n$ [cycles/sec] \\

\hline
$\mathcal{A}_k(t)$ & Set of active public WSs in node $k$
& $A_k(t)$ & Number of active public WSs in node $k$
& $\boldsymbol{G}$ & Connectivity matrix of ECC nodes \\

\hline
$\hat{\psi}_{n,k}^{pub}(t)$ & Computation starting time slot of $w_{n,k}^{pub}(t)$
& ${\bf L}(t)$ & Load history matrix at time slot $t$ 
& $w_d$ & Delay awareness weight \\

\hline
$W$ & Load history window [time slots]
& ${\bf l}_n^{pub}(t)$ & Lengths of public WSs hosting workloads of EA $n$
& $w_e$ & Energy consumption awareness weight \\

\hline
$\Lambda$ & Set of public WS length values [bits]
& $\tau_n^{\text{local}}$ & Overall delay of $w_n(t)$ for local processing
& $\tau_{n,k}^{\text{horiz}}$, $\tau_{n,k}^{\text{vert}}$ & Overall delay of $w_{n,k}(t)$ for horizontal (vertical) offloading\\

\hline
$e_n(t)$ & Energy consumption of $w_n(t)$ \par [Joule]
& $e_n^{priv}(t)$ & Energy consumption for local processing of $w_n(t)$ [Joule]
& $e_{n,k}^{pub}(t)$ & Energy consumption for processing $w_n(t)$ in node $k$ [Joule] \\

\hline

\end{tabular}
\end{table*}

\subsection{General Architecture}\label{subsec:system_architecture}

\begin{figure}[t]
\centering
\includegraphics[trim={0.2cm 0.8cm 0.2cm 0.3cm},clip,width=1\columnwidth]{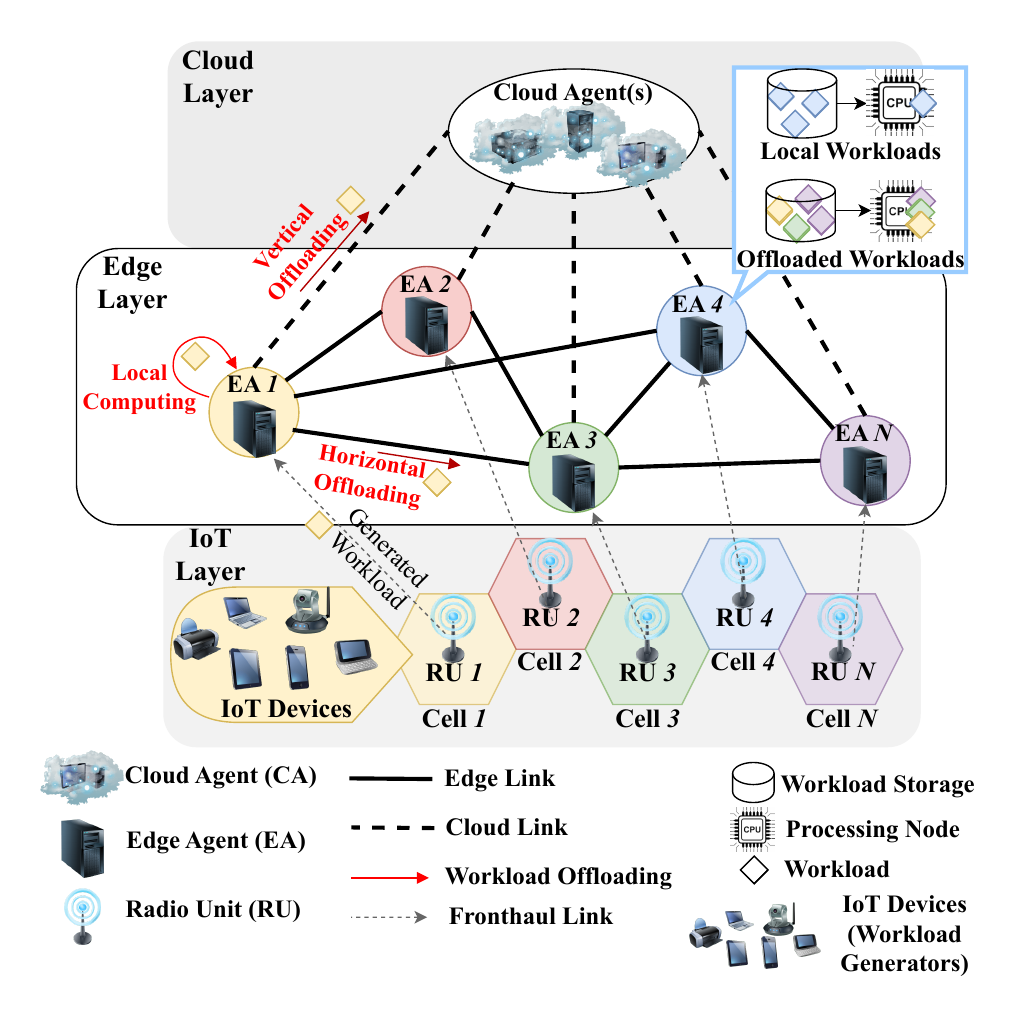}
\caption{Three-tier computing continuum architecture. Edge Agents (EAs) receive workloads from the multi-cell IoT layer (served by Radio Units) and decide between three actions: local computation, horizontal offloading to another EA, vertical offloading to the Cloud Agent (CA).}
\label{fig1}
\end{figure}

As illustrated in Fig.~\ref{fig1}, we consider a three-tier architecture comprising IoT devices, Edge Agents (EAs), and a Cloud layer, supporting both edge and cloud computing services. IoT devices of each cell are served by a Radio Unit (RU), which is associated with an EA for edge computing. Thus, to host the computational workload generated by the IoT cells, the computing continuum includes $N$ EAs and a single Cloud Agent (CA), without loss of generality. We define the set of all computing nodes as $\mathcal{N}'=\{1,2,\dots,N,N+1\}$, where node $N+1$ refers to the CA. The subset $\mathcal{N}=\{1,2,\dots,N\}$ contains only the indices of EAs. Each EA $n \in \mathcal{N}$ handles the workload generated by its associated IoT cell. Workloads are assumed to be atomic and non-partitionable, meaning they must be fully executed by a single EA or CA \cite{tang2020deep}. Given the expected high volume of workload requests in dense, 6G-enabled environments, efficient offloading decisions are essential to maintain performance.

To support flexible workload management, each EA in the system can choose between three actions: (i) compute the workload locally using its own computational resources, (ii) offload it horizontally to another EA, or (iii) offload it vertically to the CA \cite{giannopoulos2024cooler}. Each workload is time-critical and is associated with a deadline constraint, representing the maximum allowable delay until processing completion. To enable decentralized intelligence, each EA is assisted by a local DRL agent to make offloading decisions based on local observations (EA state, workload information) and shared telemetry information from the computing continuum nodes. The objective is to minimize a composite cost function that accounts for workload delay, workload drop ratio, and energy consumption.

In line with computing continuum design principles \cite{fida2023bottleneck}, we further assume the presence of $M$ Telemetry Agents (TAs) distributed across the edge layer. Each TA monitors either a single ($M=N$) or a group of EAs ($M<N$), collecting telemetry or performance metrics such as load, Central Processing Unit (CPU) utilization, energy usage, workload status. TAs are also connected with each other to form a telemetry continuum, enabling data sharing across clusters for improved system awareness and coordination \cite{fida2023bottleneck, otero2024towards}. For the formulation, we consider a single episode consisting of a finite time horizon $\mathcal{T}=\{1,2,\dots,T\}$. Also, each time slot
$t \in \mathcal{T}$ has a fixed duration of $\Delta$ seconds.

Regarding communication channels, each IoT cell is served by a dedicated RU that connects to nearby IoT devices via wireless radio access technologies such as 5G. Each RU $n \in \mathcal{N}$ is co-located or is linked to its corresponding EA $n$ through a high-speed wired backhaul (e.g., optical fiber). All EAs have access to the Internet for migrating workloads towards the CA. Connectivity between EAs is ensured through wired technologies (e.g. fiber optics) to allow peer-to-peer workload offloading \cite{giannopoulos2024pdppnet}, while the binary and symmetrical connectivity matrix $\boldsymbol{G}_{N \times N}$ defines which pairs of nodes are inter-connected. In specific, $\boldsymbol{G}_{i,j}=1$ when EA $i \in \mathcal{N}$ has wired connection to EA $j \in \mathcal{N}$.

\begin{figure}[t]
\centering
\includegraphics[trim={0 0.9cm 0 0.7cm},clip,width=1\columnwidth]{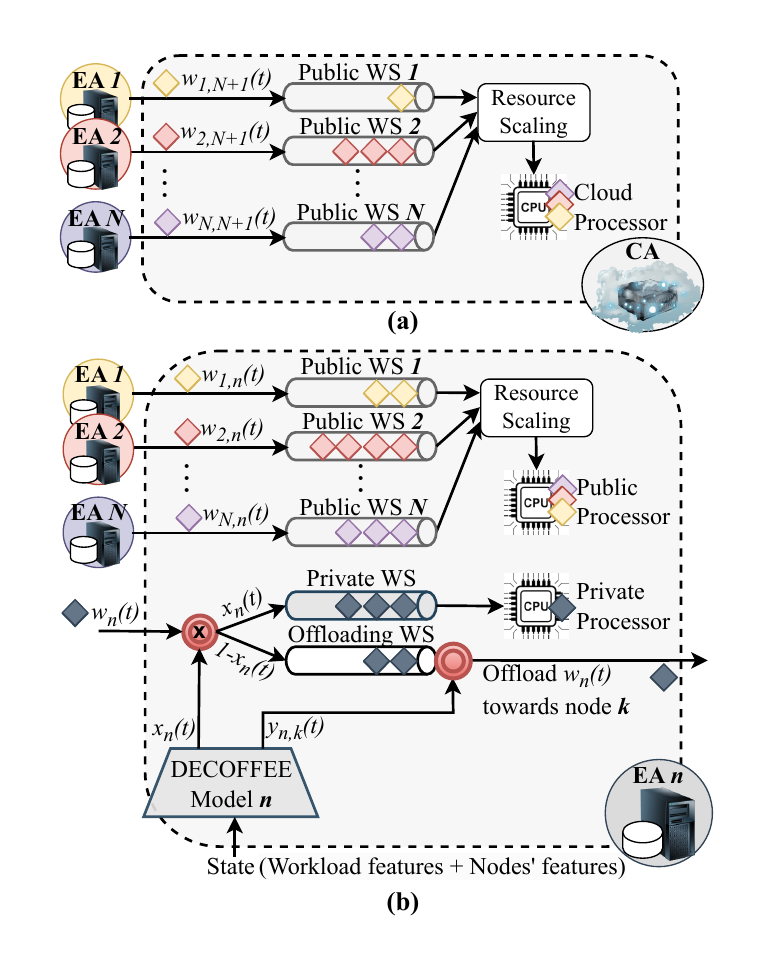}
\caption{Workload placement and storage structures inside Cloud (CA) and Edge Agents (EAs). \textbf{(a)} CA contains $N$ public workload stacks (WSs) to store workloads offloaded by EAs. \textbf{(b)} EA $n$ includes $1$ private WS to store workloads for local computation, $N-1$ public WSs for offloaded workloads and $1$ offloading WS for local workloads to be offloaded. Placement decisions are provided by the DECOFFEE model.}
\label{fig2}
\end{figure}

The internal workload placement and storage mechanisms of EAs and CA are illustrated in Fig.~\ref{fig2}. An EA may receive a new (local) workload from its associated IoT cell with probability $\mathcal{P}$ at the start of each time slot \cite{liu2016delay}. The role of each EA is two-fold: it can (i) compute workloads originating from its own IoT cell (locally), or (ii) process workloads that have been offloaded from other EAs. To support this, each EA $n \in \mathcal{N}$ maintains $N$ workload stacks (WS) for processing and one offloading WS for holding workloads awaiting offloading. All WSs operate in a First-In First-Out (FIFO) manner. The $n^{\text{th}}$ WS of EA $n \in \mathcal{N}$ is called private WS, reserved for local workloads that are decided to be processed locally. The remaining $N-1$ WSs are called public WSs, each dedicated to storing workloads offloaded by another EA. Specifically, public WS $k \in \mathcal{N} \setminus \{n\}$ of each EA $n$ stores workloads offloaded by EA $k$\footnote{This one-to-one mapping between offloading EA and destination WS avoids collisions when multiple offloads target the same EA simultaneously. Practically, each offloading EA $n$ is linked via a dedicated wired connection to public WS $n$ of the destination EA or CA.}. For example, if EA $n=10$ offloads a workload to EA $k=5$, it will be placed in the public WS $10$ of EA $5$. CA maintains only public WSs (one for each EA) where public WS $n \in \mathcal{N}$ stores workloads offloaded from EA $n$. We also assume that, upon completion of workload computation or offloading within a time slot, the next workload is scheduled for processing at the start of the following slot \cite{liu2016delay}.

\subsection{Workload Description}\label{subsec:work_feautures}

Let $w_n(t)$ represent the workload assignment number (positive integer) for the local workload received by EA $n \in \mathcal{N}$ at time slot $t \in \mathcal{T}$. The arrival index $z_n(t) \in \{0,1\}$ indicates whether a new local workload arrives at EA $n$ during slot $t$, defined as $z_n(t)=1$ (if a new workload arrives at EA $n$ at time $t$, or $z_n(t)=0$ (no workload arrival). The size of the workload  $w_n(t)$, denoted by $\eta_n(t)$ (in bits), is sampled from a discrete set $\mathcal{H} = \{0, \eta_1, \eta_2, \dots, \eta_{|\mathcal{H}|}\}$. By convention, we assume $w_n(t)=0$ and $\eta_n(t)=0$, when $z_n(t)=0$. Each workload $w_n(t)$ is also associated with a processing density $\rho_n(t)$ (in CPU cycles per bit), and a deadline $T_n^{\text{max}}(t)$, representing the number of time slots available for execution. This implies that the workload $w_n(t)$ must be completed by slot $t + T_n^{\text{max}}(t) - 1$; otherwise, it is dropped due to deadline violation.

\subsection{Workload Placement Decision}\label{subsec:place_decision}

When a new workload $w_n(t)$ arrives in EA $n \in \mathcal{N}$, a two-level placement decision is triggered to determine its execution destination. The workload placement decision is represented by the triplet $\boldsymbol{D}_n(t)=\Bigl( x_n(t), y_{n,k}(t), k \Bigr)$, where the first-level decision variable $x_n(t)$ determines local computing or offloading, while the second-level decision variable $y_{n,k}(t)$ specifies the offload destination after the workload is queued for offloading (see Fig.~\ref{fig2}). Specifically, the first-level decision $x_n(t)$ is defined as:
\begin{equation}
    x_n(t)= 
\begin{cases}
    1,& \text{if $w_n(t)$ is placed in the private WS} \\
    0,& \text{if $w_n(t)$ is placed in the offloading WS}
\end{cases}
\end{equation}

\noindent If $x_n(t) = 0$ (i.e., workload $w_n(t)$ is to be offloaded), then $y_{n,k}(t)$ decides the destination node $k \in \mathcal{N}' \setminus \{n\}$, which is formally set as:
\begin{equation}
    y_{n,k}(t)= 
\begin{cases}
    1,& \text{if $w_n(t)$ is offloaded from EA $n$ to node $k$} \\
    0,& \text{if $x_n(t) = 1$ (i.e., local execution)}
\end{cases}
\end{equation}

\noindent Here, $y_{n,k}(t)$ indicates that the workload is transferred from the offloading WS of EA $n$ to the public WS $n$ of node $k \in \mathcal{N}' \setminus \{n\}$ (offloading to the originating EA itself is not allowed), which can be another EA or the CA. For instance, when $\boldsymbol{D}_5(t)=\Bigl( 1, 0, 0 \Bigr)$, this means that workload $w_5(t)$ will be placed in the private WS for local processing, whereas a $\boldsymbol{D}_5(t)=\Bigl( 0, 1, 7 \Bigr)$ indicates that workload $w_5(t)$ will be offloaded to the public WS of node $7$.

As a result, the number of bits queued in the private and offloading WS of EA $n$ at time $t$ is given by $x_n(t) \cdot \eta_n(t)$ and $(1 - x_n(t)) \cdot \eta_n(t)$, respectively. To enforce that each workload is offloaded to at most one destination, we apply the following constraint:

\begin{equation}\label{ynk}
    \sum_{k \in \mathcal{N}' \setminus \{n\}} y_{n,k}(t) \leq 1, \quad \forall n \in \mathcal{N}
\end{equation}

\subsection{Local Workload Computation}\label{subsec:local_comp}

Each workload $w_n(t)$ designated for local execution is placed in the private WS of EA $n \in \mathcal{N}$. These workloads are processed using a CPU with constant processing capacity $f_{n}^{EA,priv}$ (in CPU cycles per second), which is dedicated only to local workloads for fast processing\footnote{It could be safely assumed that other types of computational power are also available for workload execution, such as Graphics Processing Unit (GPU), Digital Signal Processor (DSP), etc.}. 

Assuming that a workload $w_n(t)$ is placed in the private WS at time slot $t \in \mathcal{T}$, we denote the corresponding completion time slot by $\psi_n^{priv}(t)$. If no workload is placed at that time, we set $\psi_n^{priv}(t) = 0$. Completion refers either to the successful processing of the workload before its deadline or to its expiration (drop) if the deadline is missed. To compute this, we first define the waiting time $\tau_n^{priv}(t)$, representing the number of slots the workload remains in the private WS before its execution begins. If a workload is scheduled for local processing, it remains in the private WS for $\tau_n^{priv}(t)$ slots\footnote{The value $\tau_n^{priv}(t)$ is estimated by EA $n$ before placing $w_n(t)$ in the private WS.}, which is given by the following formula:

\begin{equation}\label{taupriv}
    \tau_n^{priv}(t) = \max\left\{ 0,\max_{t'<t}\{ \psi_n^{priv}(t') \} - t + 1 \right\}
\end{equation}

\noindent The outer $\max$ operator ensures non-negativity, while the inner $\max$ operator finds the latest completion time among previously submitted workloads, yielding the total wait before processing begins. Based on this, the completion time slot of workload $w_n(t)$ is given by:

\begin{equation}\label{psipriv}
\begin{split}
    \psi_n^{priv}(t) & = \min \Biggl\{ t+\tau_n^{priv}(t)+ \Biggl\lceil \frac{\eta_n(t) \cdot \rho_n(t)}{f_{n}^{EA,priv} \cdot \Delta} \Biggr\rceil -1, \\
    & \quad t + T_n^{\text{max}} - 1 \Biggr\}
\end{split}
\end{equation}

\noindent The first term corresponds to the computed finish time based on workload size, density, CPU rate, and slot duration \( \Delta \). The second term reflects the workload’s deadline. If the required waiting and processing time exceed the deadline, the workload is dropped. For example, assume a new workload $w_3(5)$ arrives at EA $3$ in time slot $5$ and is locally executed. The previous workloads $w_3(1)$, $w_3(2)$, $w_3(3)$, and $w_3(4)$ completed at slots $\psi_3^{priv}(1)=3$, $\psi_3^{priv}(2)=4$, $\psi_3^{priv}(3)=5$, and $\psi_3^{priv}(4)=12$, respectively. The waiting time for workload $w_3(5)$ is calculated as $
\tau_3^{priv}(5) = \max\left\{ 0, \max\{3, 4, 5, 12\} - 5 + 1 \right\} = 8$. Thus, workload $w_3(5)$ must wait $8$ time slots before execution begins.

\subsection{Local Workload Offloading}\label{subsec:offloading_queue}

To manage local workloads that are scheduled for offloading, each EA maintains a FIFO offloading WS. If a local workload $w_n(t)$ of EA $n \in \mathcal{N}$ is selected for offloading, it is transferred from the offloading WS to the corresponding public WS of the target EA (or the CA) via a dedicated wired link\footnote{For simplicity, we assume wired interconnections, but this can be generalized.}. Given that full connectivity between all EAs may not always be feasible, the symmetric adjacency matrix $\boldsymbol{G}$ describes the Edge layer connectivity\footnote{While a time-varying $\boldsymbol{G}(t)$ can be easily adopted to model dynamic topologies, here we assume it remains fixed.}. In a fully connected scenario, all elements of $\boldsymbol{G}$ equal 1. The transfer data rate (in Mbps) for EA-to-EA (horizontal) and EA-to-CA (vertical) communication is denoted by $R_{n,k}^H$ (connection between EA $n$ and $k$) and $R_{n,N+1}^V$ (connection between EA $n$ and CA), respectively. Transfer data rates satisfy $R_{n,k}^H > R_{n,N+1}^V$ for all $(n,k)$ horizontal (EA-to-EA) connections and all $(n,N+1)$ vertical (EA-to-CA) connections. The set of all data rates is $\mathcal{R} = \mathcal{R}_H \cup \mathcal{R}_V$, where the sets $\mathcal{R}_H$ and $\mathcal{R}_V$ contain the available data rates for horizontal and vertical links, respectively. When EA \( n \) offloads \( w_n(t) \) to destination node \( k \in \mathcal{N}' \setminus \{n\} \), the effective offloading rate is:

\begin{equation}
    R_{n,k}= 
\begin{cases}
    R_{n,k}^H,& \text{if } k \in \mathcal{N} \setminus \{n\} \\
    R_{n,k}^V,& \text{if } k = N+1 
\end{cases}, \quad \forall n \in \mathcal{N}
\end{equation}

\noindent If a workload $w_n(t)$ is placed in the offloading WS of EA $n$ at time slot $t$, the waiting time in the offloading WS, denoted $\tau_n^{off}(t)$, is computed as:

\begin{equation}\label{tauoff}
    \tau_n^{off}(t) = \max\left\{ 0,\max_{t'<t}\{ \psi_n^{off}(t') \} - t + 1 \right\}
\end{equation}

\noindent where $\psi_n^{off}(t)$ is the offload completion time of a workload placed in the offloading WS at time slot $t$. Evidently, the waiting time of $w_n(t)$ is computed as the positive difference between the completion time of the most time-consuming previous workload and the arrival time of $w_n(t)$. Note that the value $\tau_n^{off}(t)$ is estimated before EA $n$ decides where to place the current workload. If there is no workload placed for offloading at time $t$, we set $\psi_n^{off}(t) = 0$. The offload completion time slot of $w_n(t)$ is then given by:

\begin{equation}\label{psioff}
\begin{split}
    \psi_n^{off}(t) & = \min \Biggl\{ t + \tau_n^{off}(t) + \Biggl\lceil \sum_{k \in \mathcal{N}' \setminus \{n\}} \frac{y_{n,k}(t) \cdot \eta_n(t)}{R_{n,k} \cdot \Delta} \Biggr\rceil - \\ & - 1, \quad t + T_n^{\text{max}} - 1 \Biggr\}
\end{split}
\end{equation}

\noindent where $y_{n,k}(t) \in \{0,1\}$ indicates whether EA $n$ offloads to node $k$. The two arguments in the $\min\{\cdot\}$ operator reflect that the offload completion time $\psi_n^{off}(t)$ is either equal to the time slot of successful workload transfer (first term), or to the time slot of workload deadline exceedance (second term). If the offloading target is the CA (i.e. $y_{n,N+1}(t) = 1$), then $R_{n,N+1} = R_{n,N+1}^V$. Otherwise, if the offloading target is another EA, then $R_{n,k} = R_{n,k}^H$.

\subsection{Offloaded Workload Computation}\label{subsec:public_queue}

To support horizontal offloading across the continuum, each EA $n \in \mathcal{N}$ maintains $N-1$ public WSs to handle external workload requests offloaded by other EAs, while the CA hosts $N$ public WSs to serve offloads from all EAs. Each public WS is uniquely associated with its source EA, meaning that public WS $i$ at node $k \in \mathcal{N}'$ is designated to store workloads offloaded from EA $i \in \mathcal{N}$. When a workload from EA $n$ is delivered to node $k \in \mathcal{N'} \setminus \{n\}$ at time slot $t \in \mathcal{T}$, it is placed in public WS $n$ of node $k$ in the subsequent slot $t+1$. At this point, a new unique workload assignment number $w_{n,k}^{\text{pub}}(t) \in \mathbb{Z}_+$ is generated, ensuring traceability across the network. This assignment follows:

\begin{equation}
    w_{n,k}^{\text{pub}}(t)= 
\begin{cases}
    w_n(t'),& \text{if EA \( n \) offloads to node \( k \)}\\
    0,& \text{otherwise} 
\end{cases}
\end{equation}

\noindent where $w_n(t')$ is the original workload number received by EA $n$ at arrival time $t' \leq t$\footnote{Practically, this reassignment ensures that the identifier remains the same as the original one, but the timestamp and the offloading source-destination pair are updated.}. Let $\eta_{n,k}^{\text{pub}}(t) \in \mathcal{H} \cup \{0\}$ be the size (in bits) of the workload placed in public WS $n$ of node $k \in \mathcal{N'} \setminus \{n\}$ at time $t$, and let $l_{n,k}^{\text{pub}}(t)$ denote the length of this WS at the end of time $t$.

A public WS $n$ at node $k$ is considered active at time $t$ if either a new workload is inserted at that slot or there were residual workloads at $t-1$. The active WS set $\mathcal{A}_k(t)$ (with element count $A_k(t)$) at node $k \in \mathcal{N'}$ is:

\begin{equation}
    \mathcal{A}_k(t)= 
\begin{cases}
    \Bigl\{ n\,|\, \eta_{n,k}^{\text{pub}}(t)>0 \text{ or } l_{n,k}^{\text{pub}}(t-1)>0, \dots\\n \in \mathcal{N} \setminus \{k\} \Bigr\}, \quad \text{if } k \in \mathcal{N} \\
    \Bigl\{ n\,|\, \eta_{n,k}^{\text{pub}}(t)>0 \text{ or } l_{n,k}^{\text{pub}}(t-1)>0, \dots\\n \in \mathcal{N} \Bigr\}, \quad \text{if } k = N+1
\end{cases}
\end{equation}

\noindent where the upper branch is the set of active WSs of EAs, whereas the lower branch is the same for the CA. The public processor at each node $k \in \mathcal{N}'$, with computational speed $f_k^{\text{pub}}$, is equally allocated among the workloads of active WSs. The CA is equipped with a higher-capacity processor $f_{N+1}^{\text{pub}}$ than EAs, satisfying $f_{N+1}^{\text{pub}} > f_k^{\text{pub}}$, $\forall k \in \mathcal{N}$. Hence, the effective processing rate per active WS at time $t$ is $f_k^{\text{pub}} / A_k(t)$. This fair distribution of the processing capacity across the active public queues is complaint with the principle of the generalized processor sharing model \cite{zhang1995statistical, tang2020deep}. Thus, the processing capacity assigned to the offloaded workloads dynamically depends on the number of active WSs at each time slot and cannot be known a priori. Instead, each EA is only knows the public CPU processing capacity of the other EAs and CA. Note that prioritized processor capacity allocation policies can be flexibly assumed, without loss of generality \cite{giannopoulos2024pdppnet}. For instance, when some  EAs or workloads have higher priority than others, we can adopt a weighted processor capacity scaling based on different weights per EA or workload \cite{chen2019toffee}.

The length of the public WSs is dynamically updated based on incoming and outgoing number of bits inserted or dropped from the WS. We let $m_{n,k}^{\text{pub}}(t)$ denote the number of bits dropped from public WS $n$ of node $k$ at the end of time slot $t$. The length of the WS is updated via:

\begin{equation}\label{lengthWS}
\begin{split}
    l_{n,k}^{\text{pub}}(t) = \max \left\{ 0, l_{n,k}^{\text{pub}}(t-1) + \eta_{n,k}^{\text{pub}}(t) - m_{n,k}^{\text{pub}}(t) \right.\\
    \left. - \frac{\Delta \cdot f_k^{\text{pub}}}{\rho_n(t) \cdot A_k(t)} \right\}
\end{split}
\end{equation}

\noindent where it is implied that the length $l_{n,k}^{\text{pub}}(t)$ is the difference between the bits maintained in the public WS and the bits left the public WS. Specifically, the first two (positive) terms are the previous WS length and the newly added bits at time slot $t$, respectively. The last two (negative) terms are the bits dropped from the public WS and the bits processed at the time slot $t$, respectively. Note that, for $k = N+1$, the computations in \eqref{lengthWS} refer to the length of the public WSs at the CA.

The completion time slot $\psi_{n,k}^{\text{pub}}(t) \in \mathcal{T}$ of workload $w_{n,k}^{\text{pub}}(t)$ cannot be known a priori, since the number of offloaded workloads at public WSs is dynamic over time. Instead, the time slot of starting the computation of workload $w_{n,k}^{\text{pub}}(t)$, denoted as $\hat{\psi}_{n,k}^{\text{pub}}(t)$, can be defined via:

\begin{equation}
    \hat{\psi}_{n,k}^{\text{pub}}(t) = \max \left\{ t, \max_{t'<t} \psi_{n,k}^{\text{pub}}(t')+1 \right\}
\end{equation}

\noindent where it is implied that the computation of the offloaded workload $w_{n,k}^{\text{pub}}(t)$ starts either right after its arrival time slot $t$ or after the completion of the most delayed previous workload. Since the processing of workload is performed within the time slots $\tau \in [\hat{\psi}_{n,k}^{\text{pub}}(t), \psi_{n,k}^{\text{pub}}(t)]$, the workload size is bounded as:

\begin{equation}\label{ineq}
    \sum_{\tau=\hat{\psi}_{n,k}^{\text{pub}}(t)}^{\psi_{n,k}^{\text{pub}}(t)-1} \frac{\Delta \cdot f_k^{\text{pub}}}{\rho_n(t) \cdot A_k(\tau)} < \eta_{n,k}^{\text{pub}}(t) \leq \sum_{\tau=\hat{\psi}_{n,k}^{\text{pub}}(t)}^{\psi_{n,k}^{\text{pub}}(t)} \frac{\Delta \cdot f_k^{\text{pub}}}{\rho_n(t) \cdot A_k(\tau)}
\end{equation}

\noindent where $A_k(\tau)$ is the time-varying number of active WSs at time slot $\tau$. The left part of the double inequality ensures that there are remaining bits to be computed within the time interval $[\psi_{n,k}^{\text{pub}}(t)-1, \psi_{n,k}^{\text{pub}}(t)]$. The right part reflects that the workload size cannot be greater than the total number of bits processed within the time interval $[\hat{\psi}_{n,k}^{\text{pub}}(t), \psi_{n,k}^{\text{pub}}(t)]$.

\subsection{Delay Components and Energy Consumption Model}

\subsubsection{Delay Components}

\begin{figure}[t]
\centering
\includegraphics[trim={0 0.7cm 0 0.5cm},clip,width=1\columnwidth]{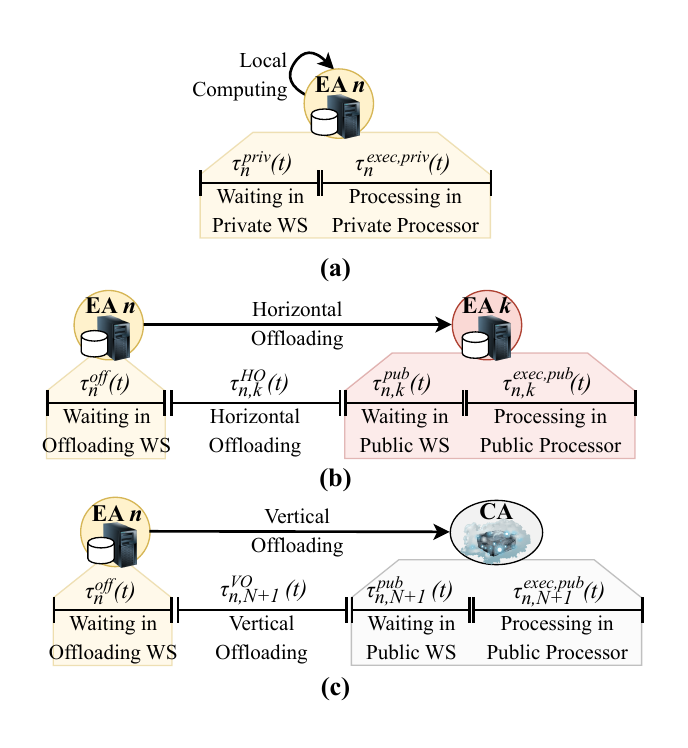}
\caption{Delay components experiences by a workload $w_n(t)$. \textbf{(a)} The case of local computation. \textbf{(b)} The case of horizontal offloading towards EA $k$. \textbf{(c)} The case of vertical offloading towards the CA.}
\label{figdelays}
\end{figure}

The delay experienced by a given workload $w_n(t)$ encompasses several components depending on the processing destination. In the considered continuum model, three distinct cases are identified, as shown in Fig. \ref{figdelays}(a)-(c). If the decision is local processing (i.e., $x_n(t)=1$), the overall delay $\tau_n^{\text{local}}(t)$ (in time slots) experienced by workload $w_n(t)$ is expressed as:

\begin{equation}
    \tau_n^{\text{local}}(t) = \min\Bigl\{ \tau_n^{priv}(t) + \tau_n^{exec,priv}(t) , T_n^{\max} \Bigr\}
\end{equation}

\noindent where $\tau_n^{priv}(t)$ is the waiting delay in the private WS (defined in \eqref{taupriv}) and $\tau_n^{exec,priv}(t)$ is the processing delay using the private processor, as defined below:

\begin{equation}
    \tau_n^{exec,priv}(t) = \Biggl\lceil \frac{\eta_n(t) \cdot \rho_n(t)}{f_{n}^{EA,priv} \cdot \Delta} \Biggr\rceil
\end{equation}

If the workload $w_n(t)$ is to be offloaded horizontally to node $k \in \mathcal{N}'$ (i.e., $y_{n,k}(t)=1$), the overall experienced delay $\tau_n^{\text{horiz}}(t)$ is the following:

\begin{equation}
\begin{split}
    \tau_{n,k}^{\text{horiz}}(t) & = \min\Bigl\{ \tau_n^{off}(t) + \tau_{n,k}^{HO}(t) + \tau_{n,k}^{pub}(t) + \\& + \tau_{n,k}^{exec,pub}(t) , T_n^{\max} \Bigr\}
\end{split}
\end{equation}

\noindent where $\tau_n^{off}(t)$ is the waiting delay in the offloading WS (defined in \eqref{tauoff}). The horizontal transfer delay $\tau_{n,k}^{HO}(t)$ to offload from EA $n$ to EA $k \in \mathcal{N}$ is:

\begin{equation}
    \tau_{n,k}^{HO}(t) = \Biggl\lceil \sum_{k \in \mathcal{N}' \setminus \{n\}} \frac{y_{n,k}(t) \cdot \eta_n(t)}{R_{n,k} \cdot \Delta} \Biggr\rceil
\end{equation}

\noindent The waiting delay in the public WS is $\tau_{n,k}^{pub}(t) = \hat{\psi}_{n,k}^{\text{pub}}(t) - t$. Also, the processing delay of the horizontally offloaded workload $w_n(t)$ using the public processor of EA $k \in \mathcal{N}$ is $\tau_{n,k}^{exec,pub}(t) = \psi_{n,k}^{\text{pub}}(t) - \hat{\psi}_{n,k}^{\text{pub}}(t)$. 

Finally, if the workload $w_n(t)$ is to be offloaded vertically (i.e., $y_{n,N+1}=1$), then the computations to find the overall delay $\tau_{n,k}^{\text{vert}}(t)$ are identical to those presented for the horizontal offloading by setting $k=N+1$ (i.e., processing destination is the CA). This affects the computations to account for the data rates $R_{n,N+1}^{V} \in \mathcal{R_V}$ of the vertical workload transfer, as well as the (public) processing capacity $f_{N+1}^{pub}$ of the CA.
All delay calculations are measured in time slots.

\subsubsection{Enegy Consumption Components}

The energy consumption for completing a workload $w_n(t)$ is measured in Joules and is computed as the product of power consumption (in Watts) and the execution duration (in seconds). Depending on the processing decision, the total energy required to complete $w_n(t)$ (from its arrival to its execution) can be calculated under two distinct cases:

\begin{itemize}
    \item If $w_n(t)$ is locally processed, the total energy consumption $e_n^{priv}(t)$ is given by:


\begin{equation}\label{energy1}
e_n^{priv}(t) =
\begin{cases}
    \Delta \cdot \Bigl( p_n^{priv}(t) \cdot T_n^{\max} \Bigr), &\\ \quad \text{if } T_n^{\max} < \tau_n^{priv}(t) \\
    \Delta \cdot \Bigl( p_n^{priv}(t) \cdot \tau_n^{priv}(t) + \dots &\\ + p_n^{exec}(t) \cdot [T_n^{\max} - \tau_n^{priv}(t)] \Bigr), &\\ \quad \text{if } T_n^{\max} < \tau_n^{priv}(t) + \tau_n^{exec,priv}(t) \\
    \Delta \cdot \Bigl( p_n^{priv}(t) \cdot \tau_n^{priv}(t) + \dots &\\ + p_n^{exec}(t) \cdot \tau_n^{exec,priv}(t) \Bigr), &\\ \quad \text{if } T_n^{\max} \ge \tau_n^{priv}(t) + \tau_n^{exec,priv}(t)
\end{cases}
\end{equation}

    \noindent where $p_n^{priv}(t)$ is the per-slot power consumption for waiting in the private WS, and $p_n^{exec}(t)$ is the per-slot power consumption for processing the workload locally. It is evident that the total energy consumption has three cases: (i) the first case corresponds to a workload being dropped without starting processing due to early deadline, (ii) the second case refers to interrupted processing before completion, consuming partial execution energy, and (iii) the third case reflects successful local computation within the deadline.
    \item If workload $w_n(t)$ is offloaded to node $k$ for processing, the total energy consumption $e_{n,k}^{pub}(t)$ is derived as:

\begin{equation}\label{energy2}
\begin{split}
    e_{n,k}^{pub}(t) & = \Delta \cdot \Bigl( p_n^{off}(t) \cdot \tau_n^{off}(t) + \dots \\& + p_{n,k}^{tran}(t) \cdot \tau_{n,k}^{HO}(t) + \dots
    \\& + p_{n,k}^{pub}(t) \cdot \tau_{n,k}^{pub}(t) + \dots
    \\& + p_{n,k}^{exec}(t) \cdot \tau_{n,k}^{exec,pub}(t) \Bigr), \\
\end{split}
\end{equation}

    \noindent where $p_n^{off}(t)$ is the per-slot power consumption for waiting in the offloading queue, $p_n^{pub}(t)$ is the power consumption for waiting in the public WS, $p_{n,k}^{exec}(t)$ is the power consumption for processing the workload of EA $n$ in the node $k$, and $p_{n,k}^{tran}(t)$ is the power consumption for transferring the workload from the offloading queue of EA $n$ to the public queue of node $k$. Thus, in this case, the total energy consumption is the aggregated energy consumed for waiting in offloading/public WSs, transferring and processing the workload.
\end{itemize}

The overall energy consumption $e_n(t)$ for workload $w_n(t)$ is then compactly expressed as:

\begin{equation}
    e_n(t)= 
\begin{cases}
    e_n^{priv}(t),& \text{if $x_n(t)=1$} \\
    e_{n,k}^{pub}(t),& \text{if $y_{n,k}(t)=1$}\\ 
\end{cases}
\end{equation}

\subsection{Load Traceability across the Continuum}

To support informed DRL-driven decisions in DECOFEEE, each TA maintains historical traces of system load levels across the computing continuum. The instantaneous load of node $n \in \mathcal{N'}$ at time slot $t$ is quantified as the number of active public queues $A_n(t)$. Accordingly, TAs maintain a time-windowed load matrix $\mathbf{L}(t) \in \mathbb{Z}_+^{W \times (N+1)}$, which records the recent $W$-slot load profiles across all nodes. Thus, the cell $(i,j)$ of $\mathbf{L}(t)$ is defined as:

\begin{equation}
    L_{i,j}(t) = A_j(t - W + i - 1), \quad \forall i,j,
\end{equation}

\noindent where each column $j$ stores the load timeseries of node $j \in \mathcal{N'}$ and each row corresponds to a past time slot, with $i \in \{1, \dots, W\}$. Load matrix is updated at the end of every time slot by the TAs. In the distributed continuum architecture, each EA-specific DRL agent receives the columns of $\mathbf{L}(t)$ by its associated TA, thus retrieving the recent history of peer nodes. Specifically, at each time slot $t$, each EA accesses the most recent $W$ load traces of peer nodes, which are essential inputs for state representation in local DRL agents, as described in Section~\ref{modelling}.

\section{Distributed Time-Critical and Energy-Efficient Workload Offloading}\label{sec:task_offloading}

In this section, we present the core principles and mathematical formulation of the proposed DECOFFEE framework. DECOFFEE enables decentralized workload offloading across the Computing Continuum by leveraging a multi-agent DRL paradigm, where each EA operates as an autonomous learning agent. At every time slot $t$, each EA observes (i) its local environment (received workloads, local WS status) and (ii) receives telemetry from other continuum nodes (others' load status) through load traceability mechanisms \cite{lin2024online}. Based on these observations, the EA selects one of three feasible actions for each incoming workload: (i) local execution, (ii) horizontal offloading to a neighboring EA, or (iii) vertical offloading to the CA.

Using DRL, each DECOFFEE agent learns long-term optimal policies by capturing the temporal evolution of system congestion and workload patterns. To this end, each DRL agent minimizes a cumulative long-term cost that jointly accounts for three key objectives: (i) workload execution latency, (ii) energy consumption during waiting, processing and transmission of workloads, and (iii) workload drop rate due to deadline violations. This joint optimization enables DECOFFEE to provide workload-aware and energy-efficient offloading decisions while satisfying strict time constraints typical of time-critical processing scenarios.

\subsection{Distributed Workload Offloading as Parallel Markov Decision Processes}

The workload offloading process in DECOFFEE is formulated as a collection of parallel agent-specific MDPs, each governing the decision-making behavior of an EA. This formalism allows each EA to autonomously determine time-critical and energy-aware offloading actions, while collectively interacting within a shared stochastic environment. At each time slot $t \in \mathcal{T}$, every EA $n \in \mathcal{N}$ may receive new workload $w_n(t)$, and must decide on the workload destination. These decisions are made sequentially and independently, yet the environment in which they are embedded is shared and coupled across agents due to resource contention, dynamic workload patterns, and network interactions. Formally, we define a family of $N$ MDPs, one per EA, where the MDP $n$ is denoted as:

\begin{equation}
    \mathcal{M}_n = \langle \mathcal{S}_n, \mathcal{A}_n, \mathcal{P}_n, \mathcal{R}_n, \gamma \rangle
\end{equation}

\noindent where the $\mathcal{S}_n$ is the state space, $\mathcal{A}_n$ is the action space, $\mathcal{P}_n$ is the transition function, $\mathcal{R}_n$ is the reward fuction, and $\gamma$ is the discount factor. The state space captures the local observables of EA $n$ at time $t$, including current workload characteristics (size, computation intensity, deadline), as well as partial observability of the other continuum nodes. The action space is the set of all possible offloading destinations for a given $w_n(t)$. The reward function quantifies the quality of an action by penalizing execution delay, energy cost, and workload loss, while the discount factor balances short- and long-term cost minimization. The transition function defines the stochastic evolution of EA $n$'s state due to both its own action and the evolving global environment. In classic MDPs, $\mathcal{P}_n$ is the probability of moving from one state to another given an action:

\begin{equation}
    \mathcal{P}_n(s'|s,a) = \probP\Bigl(s_n(t+1)=s'|s_n(t)=s,a_n(t)=t\Bigr)
\end{equation}

\noindent where $\mathcal{P}_n(s'|s,a)$ is the probability that agent $n$ moves to state $s'$ at time $t+1$, given it was in state $s$ and took action $a$ at time $t$. However, since the full system state is not directly observable and depends on the actions of other agents, $\mathcal{P}_n$ is not known in closed form because (i) the environment (i.e., the rest of the continuum) is dynamic and only partially observable, and (ii) other agents' actions affect the state of EA $n$ indirectly (e.g., length of public WSs at a target EA). In the DECOFFEE framework, the next state may be updated as:

\begin{equation}
    s_n(t+1) = f_{\text{env}}\Bigr(s_n(t),a_n(t),\boldsymbol{\zeta}_{\text{env}}(t)\Bigl)
\end{equation}

\noindent where $f_{\text{env}}(\cdot)$ is an unknown environment transition function, $\boldsymbol{\zeta}_{\text{env}}(t)$ captures stochastic environment effects (e.g., new workload arrivals, queuing delays, public CPU contention), and
current action $a_n(t)$ affects the next state via changes in queuing delay and remaining load. Hence, we can implicitly learn the transition function $\mathcal{P}_n(s'|s,a)$ of all DECOFFEE agents via model-free DRL with experience replay, where each DRL agent approximates optimal policies by interacting with the environment over time. 

\subsection{DRL Formulation for Parallel MDPs}\label{modelling}

\begin{figure}[t]
\centering
\includegraphics[width=1\columnwidth]{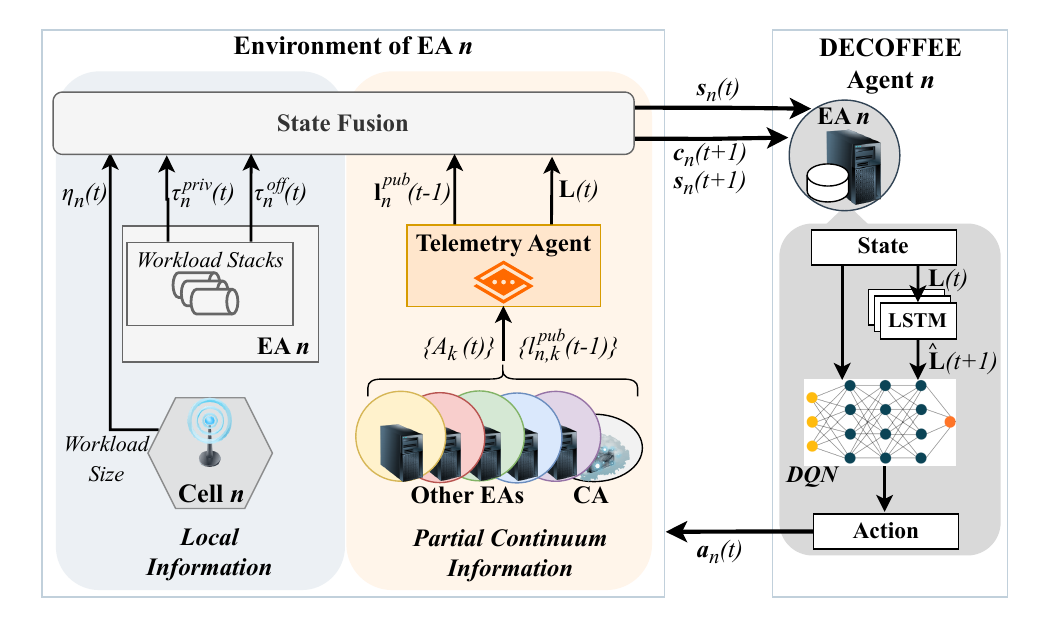}
\caption{The DRL interaction cycle between DECOFFEE agent $n$ and its observed environment. The state vector of EA $n$ combines both local information and partial observability from the other continuum nodes.}
\label{figdrl}
\end{figure}

The workload offloading problem in DECOFFEE is modeled as a collection of independent but coupled MDPs, where each EA $n \in \mathcal{N}$ operates as a distributed DRL agent interacting with its local environment, upon partially observing the global system state (see Fig.~\ref{figdrl}). Each EA in DECOFFEE is equipped with a DRL agent that interacts with its environment through three principal components: the observable state $s_n(t)$, the available action $a_n(t)$, and the corresponding cost (reward) signal $c_n(t)$. These components are formally below.

\paragraph{State Space} Let $\mathcal{S} = \{\mathcal{S}_1, \mathcal{S}_2, \dots, \mathcal{S}_N\}$ denote the collection of agent-specific state spaces, where $\mathcal{S}_n$ represents the finite and discrete set of observable states for EA $n \in \mathcal{N}$. At the beginning of each time slot $t \in \mathcal{T}$, EA $n$ observes the local state vector $s_n(t) \in \mathcal{S}_n$, defined as:

\begin{equation}
\label{state_vec}
    s_n(t) = \Bigl[ \eta_n(t), \tau_n^{priv}(t), \tau_n^{off}(t),{\bf l}_n^{pub}(t-1), {\bf \hat{L}}(t+1)]
\end{equation}

\noindent where ${\bf l}_n^{pub}(t-1)$ captures the lengths of the public WSs that host offloaded workloads from EA $n$ to other nodes, and ${\bf \hat{L}}(t+1)$ is the $N$-size vector of LSTM-predicted load values for all $N$ nodes at the next time slot $t+1$. Specifically, ${\bf l}_n^{pub}(t-1) = \{ l_{n,k}^{pub}(t-1)\}, \forall k \in \mathcal{N'} \setminus \{n\}$. The inclusion of predicted load values ${\bf \hat{L}}(t+1)$ in the state space of EA $n$ is essential to enable proactive decision-making, because relying solely on instantaneous load observations would result in outdated decisions, as the system state may change before the action takes effect\footnote{By incorporating short-term load forecasts, each agent anticipates future congestion and selects actions that remain efficient under the next-state system conditions, compensating for the action implementation latency.}. Each EA receives shared knowledge from the respective TA, which can reconstruct ${\bf l}_n^{pub}(t-1)$ by tracking (i) the number of bits offloaded to other nodes, and (ii) the number of bits processed or discarded by each remote public WS. Also, the load $A_k(t)$ of each other node $k \in \mathcal{N'} \setminus \{n\}$ is shared by the TA.

The cardinality of the state space grows with the number of nodes, the available workload sizes, and the available WS capacity. Specifically, the size of the state space is upper bounded by $|\mathcal{S}_n| = \mathcal{H} \cdot |\mathcal{T}|^2 \cdot |\Lambda|^N \cdot |\{0,1,\dots,N\}|^{(N+1)}$, where $\Lambda$ is set of public WS length values, respectively.

\paragraph{Action Space} Upon the arrival of a new workload $w_n(t)$ at EA $n$ during time slot $t$, the DRL agent selects an action $a_n(t)$ that dictates the offloading strategy. This action $a_n(t)$ is expressed as:

\begin{equation}
\label{action_vec}
    a_n(t) = \boldsymbol{D}_n(t) = \Bigl( x_n(t), y_{n,k}(t), k \Bigr)
\end{equation}

A change in action $a_n(t)$ modifies the system state in the subsequent time step, producing a new state $s_n(t+1) \ne s_n(t)$, due to dynamic effects on private and public WSs and workload migration. The size of the action space per EA is $4N$, as the first two elements in $\boldsymbol{D}_n(t)$ are binary and the third takes $N$ values.

\paragraph{Cost Function} The cost incurred by EA $n$ upon taking action $a_n(t)$ at state $s_n(t)$ is indirectly encoded via a reward signal reflecting the cost function $c_n(t')$, received at a future time slot $t'>t$ when the workload is resolved (either dropped or processed). The cost function is defined as:
 Here we consider the joint minimization of the workload execution delay (i.e. the time interval between workload arrival and workload execution), the workload energy consumption (i.e. the energy consumed from workload arrival to execution) and the workload loss ratio (due to timeout violation). To this end, the reward $r_n(t+1)$ is given by:

\begin{equation}\label{reward}
    c_n(t'>t)= 
\begin{cases}
    \emptyset,&\text{if $x_n(t)=0$ (no arrival)}\\ \Phi_n(t),&\text{if $\psi_n^{priv}(t)<t+\phi_n-1$}\\ &\text{or $\psi_{n,k}^{pub}(t')<t+\phi_n-1$}\\
    &\text{(processed)}\\
    C,& \text{otherwise (dropped)} 
\end{cases}
\end{equation}

\noindent where the cost $\Phi_n(t)$ received for workload $w_n(t)$ is a weighted sum of the execution delay and the energy consumed from arrival to completion of workload $w_n(t)$:

\begin{equation}
\Phi_n(t) = 
\begin{cases}
\Phi_n^{priv}(t), & \text{if } x_n(t) = 1 \\
\Phi_n^{pub}(t), & \text{if } x_n(t) = 0 \text{ and } y_{n,k}(t)=1
\end{cases}
\end{equation}

\noindent where $\Phi_n^{priv}(t)$ is the cost for processing the workload $w_n(t)$ locally and is given by:

\begin{equation}
    \Phi_n^{priv}(t) = w_d \Bigl( \psi_n^{priv}(t)-t+1 \Bigr) + w_e e_n^{priv}(t)
\end{equation}

\noindent and $\Phi_n^{pub}(t)$ is the cost for processing the workload in another node and is computed as:

\begin{equation}\label{costpub}
\begin{split}
    \Phi_n^{pub}(t) = & \sum_{k \in \mathcal{N} \setminus \{n\}} \sum_{t'=t}^T y_{n,k}(t) \Bigl[ 
    w_d \Bigl( \psi_{n,k}^{pub}(t')- \\& t+1 \Bigr) + w_e e_{n,k}^{pub}(t') \Bigr], \forall n\in\e
\end{split}
\end{equation} 
    
Importantly, both cases of the workload processing cost is a weighted sum of the workload execution delay (first term) and the workload energy consumption (second term), with $w_d$ and $w_e$ being constant scalars that regulate the contribution of the delay and energy consumption term, respectively\footnote{Additional terms targeting to other objectives, such as priorities or subscription fee for different classes of users charged by the nodes, can be also considered without loss of generality.}. If the execution delay exceeds the workload-specific timeout, a fixed penalty $C > 0$ is applied, increasing the agent’s sensitivity to deadline violations.

\subsection{Policy Optimization Problem}

From a system perspective, the $N$ MDPs evolve in parallel and asynchronously (i.e., EA-specific rewards are received in different slots), each governed by an autonomous agent learning its own policy $\pi_n: \mathcal{S}_n \rightarrow \mathcal{A}_n$. However, these MDPs are weakly coupled, since the action of one agent may affect the state transitions and reward structures of others. This introduces non-stationarity in each agent’s environment, a well-known challenge in distributed RL \cite{hazarika2023multi}. Nevertheless, under the assumption of asynchronous and partially observable dynamics, the decentralized offloading process can be modeled as a factored MDP with parallel agents acting under local information, leveraging continuum-wide load forecasts as a proxy for global awareness. This is particularly suitable for edge/cloud continuum systems, where real-time responsiveness and scalability necessitate lightweight (single-agent) decision-making without centralized coordination.

Each agent seeks to learn a policy $\pi_n^*$ that minimizes the expected cumulative cost:

\begin{equation}\label{objective}
\begin{split}
\pi_n^* & = \arg\min_{\pi_n} \mathbb{E}_{\pi_n} \left[ \sum_{t=0}^{\infty} \gamma^t \cdot c_n \Bigl(s_n(t), a_n(t) \Bigr) \right] \\&
    \textit{subject to: } \text{constraints \eqref{ynk}-\eqref{psioff}, \eqref{ineq},\eqref{reward}-\eqref{costpub}}
\end{split}
\end{equation}

\noindent where $\gamma \in (0,1]$ is the discount factor, and the expectation $\mathbb{E}\{ \cdot \}$ is taken over the stochastic evolution of the system, workload arrivals, and other agents' behavioral patterns, under policy $\pi_n$. To solve \eqref{objective}, DECOFFEE employs a model-free DRL framework, where each EA uses an LSTM-enhanced Deep $Q$-Network (DQN) to handle temporal dynamics, partial observability, and the inherently stochastic, non-stationary nature of the continuum.

\section{DECOFFEE Algorithm}\label{sec:decoffee}

\begin{figure}[t]
\centering
\includegraphics[width=1\columnwidth]{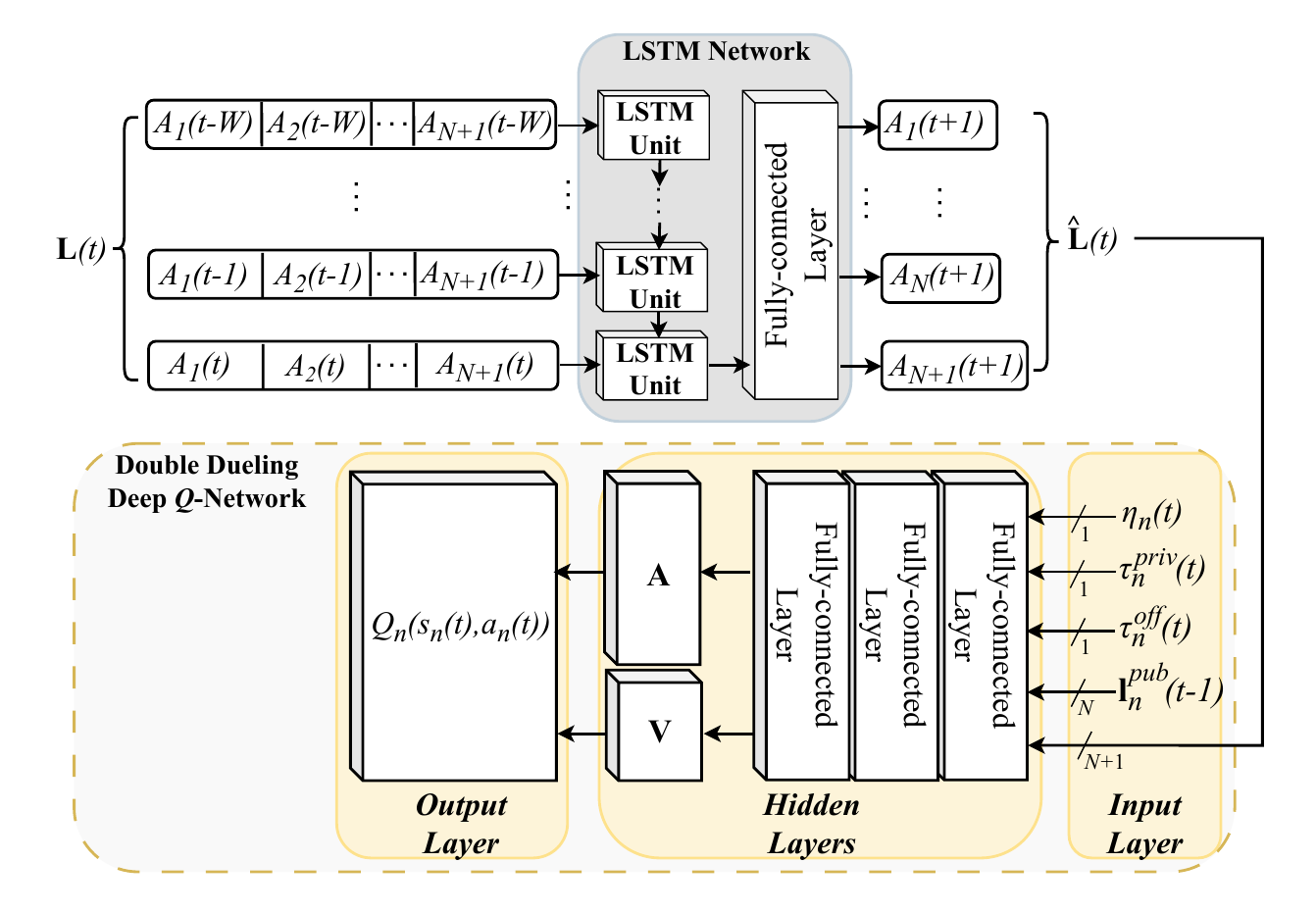}
\caption{The internal architecture of DECOFFEE agent $n$, including stacked LSTM units and a Double Dueling DQN.}
\label{fig:dqn}
\end{figure}

To address the policy optimization problem presented in \eqref{objective}, we propose DECOFFEE as a decentralized DRL-based solution tailored to the computing continuum paradigm. DECOFFEE leverages the Deep Q-Learning (DQL) framework \cite{tan2017deep}. By adopting a model-free DRL approach, DECOFFEE eliminates the need for explicit modeling of the global system dynamics or the inter-dependencies among agents, thereby offering scalability and generalizability across heterogeneous continuum environments.

Each DRL agent $n \in \mathcal{N}$ maintains a deep neural architecture, referred to as the Deep $Q$-Network (DQN), to approximate the optimal state-action value function $Q_n\bigl(s_n(t),a_n(t)\bigr)$. The $Q$-value quantifies the expected cumulative reward starting from state $s_n(t)$ and applying action $a_n(t)$, followed by the current policy thereafter. As shown in Fig.~\ref{fig:dqn}, the DECOFFEE architecture combines (i) a forecasting module based on LSTM for predictive state enrichment and (ii) a dueling DQL module for efficient policy learning.

\subsection{DECOFFEE Deep $Q$-Network}

The internal archicture of the DECOFFEE neural network model is depicted in Fig. \ref{fig:dqn}. During the training phase, the DECOFFEE model $n$ seeks to properly adjust all the network parameters $\theta_n$ (i.e. the matrix containing the weights of all connections and the biases of all neurons), such that the $Q$-values of all possible state-action pairs are accurately predicted. Below, we describe each DECOFFEE layer.

\paragraph{Input layer}

The input to the DECOFFEE model at time slot $t$ is the state vector $s_n(t)$ defined in \eqref{state_vec}, which encapsulates: the local task size $\eta_n(t)$, the waiting time in private and offloading WSs ($w_n^{priv}(t), w_n^{off}(t)$), the public WSs lengths from other nodes ${\bf l}_n^{pub}(t-1)$, and the predicted load $\hat{{\bf L}}(t)$, respectively. Notably, the predicted load vector is obtained by passing the $W$-length historical load sequences through a stack of LSTM units (top part of Fig.~\ref{fig:dqn}). This forecasting step enables each agent to anticipate future congestion\footnote{The look-ahead forecasting horizon of the LSTM is a hyperparameter that needs to be tuned for optimal DECOFFEE performance.} across the continuum nodes, rather than reacting to current information. Such proactive decision-making is critical in DECOFFEE, as workload placement decisions made at time $t$ affect system behavior in the subsequent time slots. Without prediction, the delayed execution of an action may lead to suboptimal outcomes due to system transitions that occurred after decision issuance.

\paragraph{Fully-connected hidden layers}

The predicted values $\hat{{\bf L}}(t)$ are concatenated with the rest of the input features and passed through multiple fully-connected layers. These layers constitute the backbone of the $Q$-network, enabling high-dimensional feature extraction and representation learning. Each layer contains a fixed number of hidden neurons activated via Rectified Linear Unit (ReLU) functions. As with standard DQL architectures, the weights of these layers are optimized during training to reduce the Bellman's error between predicted and target $Q$-values \cite{barron1989bellman}.

\paragraph{Dueling DQL layer}

To further improve learning stability and convergence, DECOFFEE adopts a Dueling DQN architecture \cite{giannopoulos2024pdppnet}, which decomposes the $Q$-value function into two distinct components, including (i) a state-value function $V_n(s_n(t)|\theta_n)$ that captures the overall quality of being in state $s_n(t)$, and (ii) an advantage function $A_n(s_n(t), a_n(t),|,\theta_n)$ that reflects the relative merit of taking action $a_n(t)$ in that state. These two streams are independently estimated through parallel fully-connected sub-networks and recombined to obtain the Q-value, as follows:

\begin{equation}\label{q_value}
\begin{split}
    Q_n\Bigl( {\bf s}_n(t) &, {\bf a} \Big| \theta_n \Bigr) = V_n\Bigl( {\bf s}_n(t) \Big| \theta_n \Bigr) + \Bigl[ A_n\Bigl( {\bf s}_n(t), {\bf a} \Big| \theta_n \Bigr) \\& - \frac{1}{2^{N+1}} \sum_{{\bf a'} \in \{0,1\}^{N+1}} A_n\Bigl( {\bf s}_n(t), {\bf a'} \Big| \theta_n \Bigr) \Bigr]
\end{split}
\end{equation}    

Here, $\theta_n$ denotes the trainable parameters of the neural network for agent $n$, and $2^{N+1}$ represents the count of all possible offloading actions across the $N+1$ computing nodes (i.e., $N$ EAs and the Cloud). Note that, practically, the action vector is encoded as a $N+1$-size binary vector, with all elements being at $0$ except the $i^{\text{th}}$ element which is set to $1$, denoting that the processing will be hosted at node $i \in \mathcal{N'}$.

\paragraph{Output Layer}

The final output layer of DECOFFEE yields the $Q$-values for all candidate actions $\boldsymbol{a}$ under the current input state $s_n(t)$ and the trained parameters $\theta_n$. At inference time, the optimal action $a_n^*(t)$ is selected as the one that minimizes the predicted cost, as reflected below:

\begin{equation}
    a_n^*(t) = \arg\min_{\boldsymbol{a} \in \{0,1\}^{N+1}} Q_n\Bigl( s_n(t),\boldsymbol{a} \Bigl| \theta_n \Bigr)
\end{equation}

\subsection{DECOFFEE Pseudocode}

Here we present the training and inference pseudocode of the proposed DECOFFEE algorithm. DECOFFEE comprises a set of $N$ distributed agents to independently learn to make joint energy- and delay-efficient workload placement decisions. Both training and inference are performed locally within each EA, enabling decentralized learning under dynamic workloads and system heterogeneity.

\paragraph{Reward-delayed Training Workflow} Algorithm~\ref{algorithm1} describes the DECOFFEE training procedure for a single DRL agent $n \in N$, using the experience replay technique \cite{rolnick2019experience} to improve sample efficiency and learning stability. For each workload completed at time $t$ and had been received at time $t'<t$, the experience is stored as a quadruple $\bigl( s_n(t'), a_n(t'), c_n(t), s_n(t'+1) \bigr)$ in a finite replay buffer of capacity $N_R$. The agent is trained over $N_E$ episodes, with each episode spanning $T$ time slots. To approximate the optimal placement policy $\pi_n^*$ that minimizes the cost objective in \eqref{objective}, each DECOFFEE agent maintains two neural models with identical architecture: (i) the primary $Q$-network $Q_n$ with parameters $\theta_n$, used for action selection and value prediction, and (ii) the Target $Q$-network $\hat{Q}_n$ with parameters $\hat{\theta}_n$, updated less frequently to stabilize the targets during training. The target network provides delayed versions of the $Q$-values, avoiding rapid oscillations in learning and improving convergence. Every $N_{copy}$ training episodes, the parameters of the primary network are copied to the target network, meaning $\hat{\theta}_n \gets \theta_n$.

\begin{algorithm}[t!]
\caption{Training of DECOFFEE agent $n$}\label{algorithm1}
\begin{algorithmic}[1]
\STATE \textbf{Inputs:} $T$, $\alpha_{lr}$, $\gamma$, $N_E$, $N_R$, $N_B$, $N_{copy}$, $w_d$, $w_e$
\STATE Initialize Replay Memory with $N_R$ rows
\STATE Initialize $Q_n$, $\hat{Q}_n$ models with random $\theta_n$, $\hat{\theta}_n$.
\STATE Initialize $E_{current} = 0$ and $\epsilon=1$
\FOR{each episode $j = 1, 2, \dots, N_E$}
    \STATE Initialize $s_n(1)$
    \FOR{each time step $t = 1, 2, \dots, T$}
        \STATE Generate a new workload with probability $P$
        \STATE Set workload ID $w_n(t)$
        \STATE Select action $a_n(t)$ for $w_n(t)$ based on $\epsilon$-greedy
        \STATE Observe next state $s_n(t+1)$
        \FOR{each completed $w_n(t') \in \mathcal{D}_n(t)$ \eqref{eq:previoustasks}}
            \STATE Collect cost $c_n(t)$ of $w_n(t')$ based on \eqref{reward}
            \STATE Store experience $\Bigl(  s_n(t'), a_n(t'),c_n(t),s_n(t'+1) \Bigr)$
        \ENDFOR
        \STATE Sample a random batch $B$ of $N_B$ experiences
        \FOR{each experience row $i \in B$}
            \STATE Assume row format as $\Bigl( s_{n,i},a_{n,i},c_{n,i},s'_{n,i} \Bigr)$
            \IF{$t=T$ \COMMENT{Terminal time slot}}
                \STATE Set target $y_{n,i} = c_{n,i}$
            \ELSE
                \STATE \%\textit{Double $Q$-learning}\%
                \STATE Compute target $y_{n,i} =$ 
                \STATE $c_{n,i} +\gamma \cdot \hat{Q}\Bigl(s'_{n,i},\arg\min\limits_{a'}Q\Bigl(s'_{n,i}, a' \mid \theta_n\Bigr) \mid \hat{\theta}\Bigr)$
            \ENDIF
            \STATE Compute predicted $z_{n,i} = Q_n\Bigl( s_{n,i},a_{n,i} | \theta_n \Bigr)$
        \ENDFOR
        \STATE Set target values ${\bf Y}_n^{Target}(t)=\{y_{n,i}\}, \forall i \in B$
        \STATE Set predicted values ${\bf Y}_n^{Pred}(t)=\{z_{n,i}\}, \forall i \in B$
        \STATE Update parameters $\theta_n$ to minimize loss function:
        \par MSE$\Big( {\bf Y}_n^{Target}(t), {\bf Y}_n^{Pred}(t) \Bigr)$
    \ENDFOR
    \STATE Set $E_{current} = E_{current} + 1$
    \IF{mod$(E_{current}$,$N_{copy})=0$}
        \STATE Copy weights $\hat{\theta}_n = \theta_n$
    \ENDIF
    \IF{$E_{current} \leq N_E/2$}
        \STATE Decay $\epsilon = 1 - 2(CurrentEpisode-1)/N_E$
    \ELSE
        \STATE Set $\epsilon = 0$
    \ENDIF
\ENDFOR
\STATE \textbf{Output:} Optimal policy $\pi^*_n$ for action-value function $Q_n$
\end{algorithmic}
\end{algorithm}

The training algorithm takes as input a set of learning hyperparameters, including the episode length $T$, the learning rate $\alpha_{lr}$, the discount factor $\gamma$, the number of episodes $N_E$, the size of the replay buffer $N_R$, the batch size $N_B$, the target update frequency $N_{copy}$, and the weights $w_d$, $w_e$ for delay-energy trade-off regulation. The output is the trained policy $\pi_n^*$ that maps each state to the optimal action vector. At the start of each episode, the agent initializes its state as $s_n(1)$. During each time slot $t \in [1, T]$, a new workload $w_n(t)$ may arrive from the IoT layer with probability $\mathcal{P}$. If a workload arrives, the agent selects an action $a_n(t) \in \{0,1\}^{N+1}$ using an $\epsilon$-greedy strategy. This means that the agent selects a random action with probability $\epsilon$ (exploration), or the highest-$Q$ action with probability $1-\epsilon$ (exploitation). Initially, we set to $\epsilon=1$, which is linearly decreased to $0$ over the first $N_E/2$ episodes, promoting exploration in early stages and pure exploitation thereafter.

Regardless of whether a new workload arrives, the agent always transitions to the next state $s_n(t+1)$ by re-evaluating the system status and acquiring updated LSTM-based predictions $\hat{\mathbf{L}}(t+1)$. Since workload execution (locally or remotely) may span multiple slots, the costs for previous workload decisions are delayed. Therefore, the agent constructs at time slot $t$ the set $\mathcal{D}_n(t)$ of previously offloaded or locally executed workloads that have completed exactly at $t$:

\begin{equation}\label{eq:previoustasks}
\begin{split}
    \mathcal{D}_n(t) =  \Bigl\{ w_n(t') \neq 0 \Bigr| & t'<t \text{, } \psi_n^{priv}=t \text{ or } \psi_{n,k}^{pub}=t \Bigr\}
\end{split}
\end{equation}

This reflects that each agent employs a reward-delayed collection of experiences, since transition tuples of past workloads $w_n(t'<t)$ are gathered at a future time slot $t$. Each completed workload yields a cost $c_n(t)$ according to the multi-objective metric defined in \eqref{reward}, and its experience tuple is stored in the replay buffer. A random minibatch $B$ of $N_B$ experiences is sampled from the buffer. For each experience $i \in B$, the DECOFFEE agent computes (i) the target $Q$-value $y_{n,i}$ using the Double DQL formula \cite{shokrnezhad2023double}, and the predicted $Q$-value $z_{n,i}$ using the primary $Q$-network. Then, the Mean Squared Error (MSE) between the predicted and target $Q$-values is computed over the $B$ samples and is used to update the parameters $\theta_n$. Finally, every $N_{copy}$ episodes, the target network is updated by copying $\hat{\theta}_n \gets \theta_n$. This concludes one full episode of learning for agent $n$, progressively refining the local policy $\pi_n$ until convergence.

\paragraph{Inference Workflow} 
Once training is complete, each DECOFFEE agent $n \in \mathcal{N}$ operates independently using its learned $Q_n$-model to make real-time workload placement decisions. The inference phase is executed continuously at the edge site, with no gradient updates or target network interaction involved. As outlined in Algorithm \ref{algorithm2}, when a new workload arrives at time slot $t$, the DECOFFEE agent infers the optimal action $a_n(t)$ by feeding the current state $s_n(t)$ into the trained $Q_n$-model and selecting the action that minimizes the estimated cost. The decision is purely exploitative, which means that constantly $\epsilon=0$. The result is an online inference loop by multiple autonomous agents, capable of selecting delay- and energy-aware offloading actions proactively under the computing continuum dynamics.

\begin{algorithm}[t!]
\caption{Inference of DECOFFEE agent $n$}\label{algorithm2}
\begin{algorithmic}[1]
\STATE \textbf{Input:} Trained $Q_n$-model with parameters $\theta_n$
\FOR{each inference step $t = 1, 2, \dots$}
    \STATE Check for new workload arrival
    \IF{workload $w_n(t)$ is received}
        \STATE Construct current state $s_n(t)$
        \STATE Select action $a_n(t) = \arg\min\limits_{{\bf a}} Q_n\Bigl( s_n(t), {\bf a} \mid \theta_n \Bigr)$
        \STATE Execute placement decision $a_n(t)$ for $w_n(t)$
    \ENDIF
\ENDFOR
\end{algorithmic}
\end{algorithm}

\subsection{Time Complexity}

The computational complexity of DECOFFEE can be assessed separately for the training and inference phases.

\textbf{Training Complexity:} The training procedure is episodic and unfolds over $N_E$ episodes, each consisting of $T$ time steps. For every time step $t$ within an episode, the agent may generate a new workload and compute the optimal placement decision. The experience of each workload is stored and used to update the $Q$-network using mini-batches of $N_B$ samples from a replay buffer of size $N_R$. Each time step may involve (i) iterating over a set $\mathcal{D}_n(t)$ of previously-arrived and currently-completed workloads (Line 12), and (ii) executing a batch update over $N_B$ experience tuples (Lines 17–27). The second loop dominates the computational cost, as it includes forward passes through the $Q$-network and target $Q$-network, as well as backpropagation steps. Thus, the worst-case number of iterations scales with $N_E \cdot T \cdot N_B$. Let $C_{forward}$ and $C_{backprop}$ denote the number of operations (cost) required for the feed-forward and backpropagation passes per experience sample, respectively. Then, the total computational complexity of training DECOFFEE per agent is $\mathcal{O}\Bigl( N_E \cdot T \cdot N_B \cdot (C_{forward}+C_{backprop}) \Bigr)$. The values of $C_{forward}$ and $C_{backprop}$ depend on the structure of the DQN (e.g., number of layers, neurons per layer, input/output dimensions), and are identical for all agents under homogeneous configurations. Note that the parameter synchronization between the primary and target networks (Lines 33) occurs every $N_{\text{copy}}$ episodes and incurs negligible overhead in comparison to the batch updates.

\textbf{Inference Complexity:} In the inference phase, each agent simply constructs the current state vector $s_n(t)$ upon workload arrival and selects the action with the minimum predicted cost by evaluating the $Q_n$-network (Line 6). This involves a single feed-forward pass without any backpropagation. Therefore, the per-sample inference complexity of DECOFFEE is $\mathcal{O}\Bigl( C_{forward} \Bigr)$. This lightweight inference cost makes DECOFFEE suitable for online deployment in real-time edge computing systems, where rapid decision-making is required for workload placement.

%
%
%
%
\section{Numerical Evaluation}\label{sec:experimental_results}

In this section, we numerically evaluate the performance of the proposed DECOFFEE framework in a realistic ECC setting. Our evaluation focuses on quantifying the benefits of DECOFFEE in terms of workload delay, energy consumption and reliability (inversely proportional to the drop ratio) under dynamic workloads and partial observability. First, we investigate the convergence behavior of the distributed DECOFFEE agents during training, highlighting the impact of critical DRL hyperparameters on learning efficiency and stability. We then perform sensitivity analyses with respect to key system-level parameters such as workload intensity, offloading link data rates, and computing capacities, in order to demonstrate the robustness and scalability of the DECOFFEE policies. Finally, we conduct a comparative analysis between DECOFFEE and several state-of-the-art baseline methods, evaluating their performance trade-offs under multi-metric constraints.

The simulation environment follows a realistic hardware and network setup, including computational capabilities, memory resources, and communication bandwidths. Each EA is provisioned with dual-core $5$ GHz CPUs for handling private and offloaded public workloads, while the Cloud is modeled with a high-performance multi-core $30$ GHz CPU capable of parallel execution of multiple workload queues. CPU capacity reflects the total computational throughput available at each node and is allocated dynamically across active queues using Completely Fair Scheduling (CFS) \cite{pabla2009completely}. CFS is implemented using a thread-based model, where each active WS corresponds to a separate thread scheduled uniformly over available cores. RAM capacities are set to $16$ GB for each EA and $64$ GB for the Cloud, supporting buffer management and multi-threaded execution \cite{pabla2009completely}.

\begin{table}[t]
\centering
\caption{Environment and DRL Parameters}
\label{table4}
\setlength{\tabcolsep}{3pt}
\begin{tabular}{c|c|c}
\hline
\textbf{Parameter} & \textbf{Symbol} & \textbf{Value} \\ 
\hline
Workload arrival probability & $\mathcal{P}$ & $0.7$ \\ 
\hline
Horizontal data rate & $R_{n,k}^H$ & $30$ Mbps \cite{tang2020deep}\\ 
\hline
Vertical data rate & $R_{n,k}^V$ & $10$ Mbps\\ 
\hline
Delay awareness weight & $w_d$ & $0.5$\\ 
\hline
Energy awareness weight & $w_e$ & $0.5$\\ 
\hline
Workload size & $\eta_n(t)$ & $[2,2.1,\dots,5]$ Mbits \cite{wang2017computation}\\
\hline
Workload density & $\rho_n(t)$ & $0.297$ gigacycles/Mbit \cite{wang2017computation}\\
\hline
DRL agents count & $N$ & 20\\ 
\hline
Connectivity matrix & $\boldsymbol{G}$ & See Fig.~\ref{fig:topology}\\ 
\hline
Private CPU frequency & $f_n^{EA,priv}$ & $5$ GHz\\ 
\hline
Public CPU frequency & $f_n^{pub}$ & $5$ GHz\\
\hline
Cloud CPU frequency & $f_{N+1}^{pub}$ & $30$ GHz\\
\hline
Number of Training Episodes & $N_E$ & $2000$\\
\hline
Number of Time slots & $T$ & $110$\\
\hline
Time slot duration & $\Delta$ & $0.1$ sec\\
\hline
Consumption in private WS & $p_n^{priv}$ & $0.1$ W/sec\\
\hline
Consumption in offloading WS & $p_n^{off}$ & $0.1$ W/sec\\
\hline
Consumption in public WS & $p_{n,k}^{pub}$ & $0.1$ W/sec\\
\hline
Consumption for transfer & $p_{n,k}^{tran}$ & $0.2$ W/sec\\
\hline
Consumption in private CPU & $p_n^{exec}$ & $1$ W/sec\\
\hline
Consumption in public EA CPU & $p_{n,k}^{pub}$ & $1$ W/sec\\
\hline
Consumption in Cloud CPU & $p_{n,k}^{pub}$ & $2$ W/sec\\
\hline
Workload timeout & $T_n^{\max}$ & $20$ slots (or $2$ sec)\\
\hline
DRL learning rate & $\alpha_{lr}$ & $5\cdot10^{-4}$\\
\hline
Discount factor & $\gamma$ & 0.99 \\
\hline
$Q$-network hidden layers & $N_{L}$ & $3 \times 1024$ neurons\\
\hline
$Q$-network optimizer & $Opt$ & Adam\\
\hline
Loss function & MSE & Lines 28-30 in Algorithm~\ref{algorithm1}\\
\hline
Update frequency & $N_{copy}$ & $500$ iterations\\
\hline
LSTM lookback window & $W$ & $10$ steps\\
\hline
LSTM hidden layers & $N_{\text{LSTM}}$ & $1 \times 20$ LSTM units\\
\hline
Replay Memory size & $N_R$ & $10000$ samples\\
\hline
Drop penalty constant & $C$ & 40\\
\hline
Batch size & $N_B$ & $64$ samples\\
\hline
\end{tabular}
\end{table}

Network parameters distinguish between horizontal (EA-to-EA) and vertical (EA-to-Cloud) offloading modes. Horizontal offloading is configured at $30$ Mbps, mimicking low-latency wired connections (e.g., optical fiber, Ethernet, or LAN setups), whereas vertical offloading operates at $10$ Mbps, representing more constrained public network backhauls (e.g., WAN, multi-hop Internet). These default values are representative but configurable, while DECOFFEE can be flexibly tested across various hardware profiles and link rates to accommodate different deployment scenarios.

All DRL training and inference experiments are implemented in Python 3.10, using the PyTorch framework (v2.0.1) with CUDA acceleration (v11.8). Training of the DECOFFEE agents is executed on a workstation equipped with an AMD Ryzen 7 7800X3D 8-Core Processor ($4.20$ GHz base frequency), $64$ GB of DDR5 RAM, and an NVIDIA RTX 4080 GPU. The implementation supports batched training with GPU acceleration for both the $Q$-model and target $\hat{Q}$-model networks per agent.

\subsection{Training Convergence and Learning Behavior}\label{sec:training}

To effectively train the DECOFFEE agents, a set of system and learning parameters were carefully configured, with particular emphasis on tuning the hyperparameters that critically affect the convergence and stability of the distributed DRL agents. Unless otherwise stated, the simulation and learning parameters used in this subsection are summarized in Table~\ref{table4}. DECOFFEE adopts an application-agnostic workload model, where incoming workloads are characterized solely by their size $\eta_n(t)$ and processing density $\rho_n(t)$, without assuming application-specific semantics. This abstraction allows DECOFFEE to capture a wide spectrum of practical workloads, ranging from lightweight tasks (e.g., database updates, control signaling, simple AI inference) to computation-intensive workloads (e.g., image processing or deep learning inference).

\begin{figure}[t]
\centering
\includegraphics[trim={0.4cm 0.6cm 0.4cm 0.5cm}, clip,width=1\columnwidth]{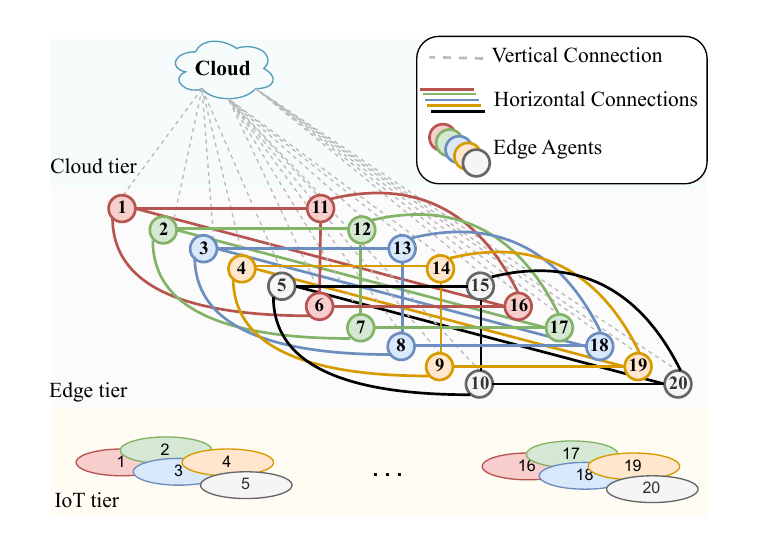}
\caption{Edge-Cloud connection topology derived from connectivity matrix $\boldsymbol{G}$.}
\label{fig:topology}
\end{figure}

Each training episode consists of $T=110$ time slots. During the first 100 slots, DECOFFEE agents actively make workload placement decisions, while the final 10 slots are reserved to drain any remaining workload stacks, ensuring that delayed rewards associated with pending workloads are properly collected. Unless explicitly stated, all subsequent evaluations follow the same configuration. In this subsection, we consider $N=20$ Edge Agents, interconnected according to the connectivity topology $\mathbf{G}$ illustrated in Fig.~\ref{fig:topology}.

\begin{table}[t]
\centering
\caption{Converged Cost for Different Hyperparameter Combinations}
\label{tab:converged_cost}
\renewcommand{\arraystretch}{1.1}
\begin{tabular}{c|c|c c c c c}
\hline
\multirow{2}{*}{\makecell{\textbf{Learning}\\\textbf{rate} ($\alpha_{\text{lr}}$)}} & 
\multirow{2}{*}{\makecell{\textbf{Discount}\\\textbf{factor} ($\gamma$)}} & 
\multicolumn{5}{c}{\textbf{Delay Coefficient} ($w_d$)} \\
\cline{3-7}
 &  & $1$ & $0.75$ & $0.5$ & $0.25$ & $0$ \\
\hline

\multirow{4}{*}{$10^{-5}$}
 & 0.25 & 0.84 & 0.99 & 1.04 & 0.97 & 1.02 \\
 & 0.50 & 0.76 & 0.95 & 0.92 & 0.88 & 0.93 \\
 & 0.75 & 0.74 & 0.91 & 0.86 & 0.81 & 0.87 \\
 & 0.99 & 0.67 & 1.12 & 0.78 & 1.13 & 1.12 \\

\hline
\multirow{4}{*}{$10^{-4}$}
 & 0.25 & 0.70 & 1.14 & 0.95 & 0.86 & 0.71 \\
 & 0.50 & 0.68 & 1.10 & 0.81 & 0.78 & 0.65 \\
 & 0.75 & 0.67 & 0.98 & 0.73 & 0.71 & 0.58 \\
 & 0.99 & 0.64 & 1.22 & 0.65 & 0.93 & 0.82 \\

\hline
\multirow{4}{*}{$5 \times 10^{-4}$}
 & 0.25 & 0.71 & 0.71 & 0.47 & 0.54 & 0.58 \\
 & 0.50 & 0.68 & 0.66 & 0.40 & 0.45 & 0.51 \\
 & 0.75 & 0.64 & \textbf{0.61} & 0.36 & \textbf{0.42} & 0.42 \\
 & 0.99 & \textbf{0.62} & 0.73 & \textbf{0.31} & 0.62 & 0.68 \\

\hline
\multirow{4}{*}{$10^{-3}$}
 & 0.25 & 0.73 & 1.25 & 0.72 & 0.64 & 0.41 \\
 & 0.50 & 0.71 & 1.20 & 0.55 & 0.53 & 0.36 \\
 & 0.75 & 0.66 & 1.12 & 0.44 & 0.49 & \textbf{0.33} \\
 & 0.99 & 0.65 & 1.32 & 0.38 & 0.71 & 0.61 \\

\hline
\end{tabular}
\end{table}

\begin{figure}[t]
\centering
\includegraphics[trim={1.5cm 0.6cm 0.2cm 0.7cm}, clip,width=1\columnwidth]{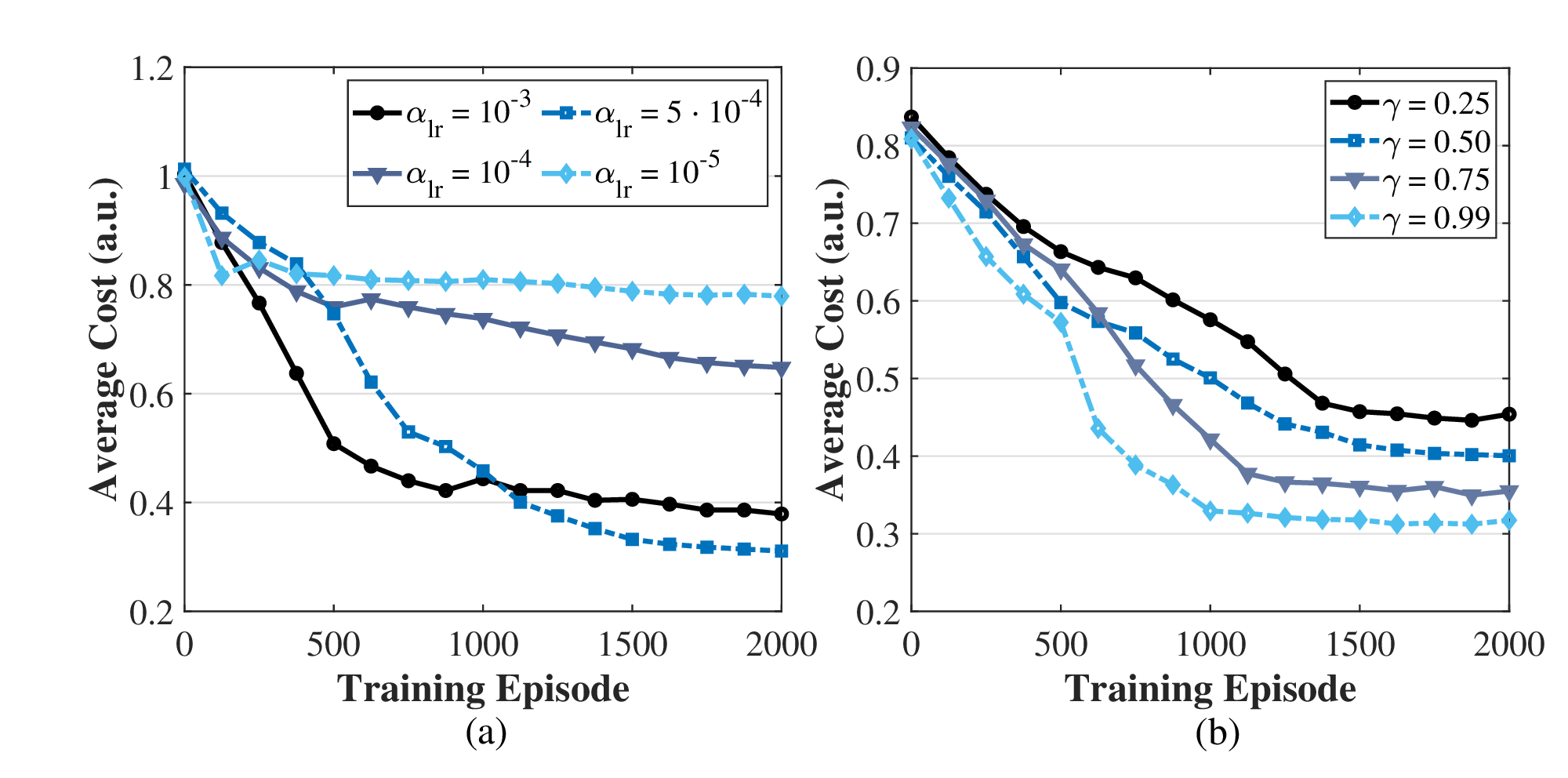}
\caption{DRL cost averaged across DECOFFEE agents as a function of training episodes for different \textbf{(a)} learning rates ($\alpha_{\text{lr}}$) and \textbf{(b)} discount factors ($\gamma$).}
\label{fig:res1}
\end{figure}

We first examine the impact of the learning rate $\alpha_{lr}$, which directly controls the magnitude of the $Q$-value updates and regulates how quickly and stably the agent updates its knowledge based on new experiences. Specifically, the learning rate was evaluated over the range $\alpha_{lr} \in \{10^{-5}, 10^{-4}, 5\cdot10^{-4}, 10^{-3}\}$, under heavy workload arrival conditions ($\mathcal{P}=0.7$). Considering equal delay and energy consumption awareness (i.e., $w_d=w_e=0.5$), Fig.~\ref{fig:res1}(a) depicts the evolution of the average cumulative cost across $2000$ training episodes, averaged over all DECOFFEE agents. As expected, excessively small learning rates result in slow convergence, while overly large values introduce instability. A learning rate of $\alpha_{lr}=5 \cdot 10^{-4}$ was found to provide the best trade-off between convergence speed and stability, yielding the lowest long-term cumulative cost. Since DECOFFEE jointly minimizes delay, energy consumption, and workload drops, the reward curves decrease over time toward zero.

Next, Fig.~\ref{fig:res1}(b) illustrates the impact of the discount factor $\gamma \in \{0.25, 0.5, 0.75, 0.99\}$. The discount factor governs the balance between immediate and future costs and plays a central role in shaping the agents’ long-term behavior. The results indicate that $\gamma=0.99$ achieves the best performance, suggesting that DECOFFEE agents benefit from strongly prioritizing long-term system efficiency. This behavior is particularly important in DECOFFEE, where offloading decisions affect not only instantaneous delay but also future energy consumption and queue dynamics across the continuum.

Table~\ref{tab:converged_cost} further complements the convergence analysis by reporting the converged long-term cost achieved by DECOFFEE under different combinations of learning rate $\alpha_{\text{lr}}$, discount factor $\gamma$, and delay-weight coefficient $w_d$. Since DECOFFEE optimizes a weighted sum of execution delay and energy consumption, varying $w_d$ directly regulates the trade-off between latency-awareness and energy-awareness. The results indicate that moderate learning rates ($\alpha_{\text{lr}}=5\times10^{-4}$) consistently yield the lowest converged cost across most values of $\gamma$ and $w_d$, confirming the observations from Fig.~\ref{fig:res1}(a). Moreover, higher discount factors ($\gamma\geq0.75$) generally lead to improved convergence, highlighting the importance of long-term planning in delay- and energy-aware workload placement. For the subsequent evaluations, we infer this wide hyperparameter space to justify the selected default configuration.

\subsection{Performance Sensitivity to Environment Parameterization}

Using the optimal hyperparameter configuration identified in Section \ref{sec:training}, the sensitivity of DECOFFEE with respect to key environment parameters is analyzed. In particular, this subsection examines how critical environment parameters affect the system behavior under different delay–energy awareness configurations. For each evaluation scenario, a series of $200$ inference episodes are executed to derive the reported metrics, using the pre-trained DECOFFEE models.

\subsubsection{Sensitivity to Workload Arrival Probability} 

\begin{figure}[t]
\centering
\includegraphics[trim={0cm 0cm 0cm 0cm}, clip,width=1\columnwidth]{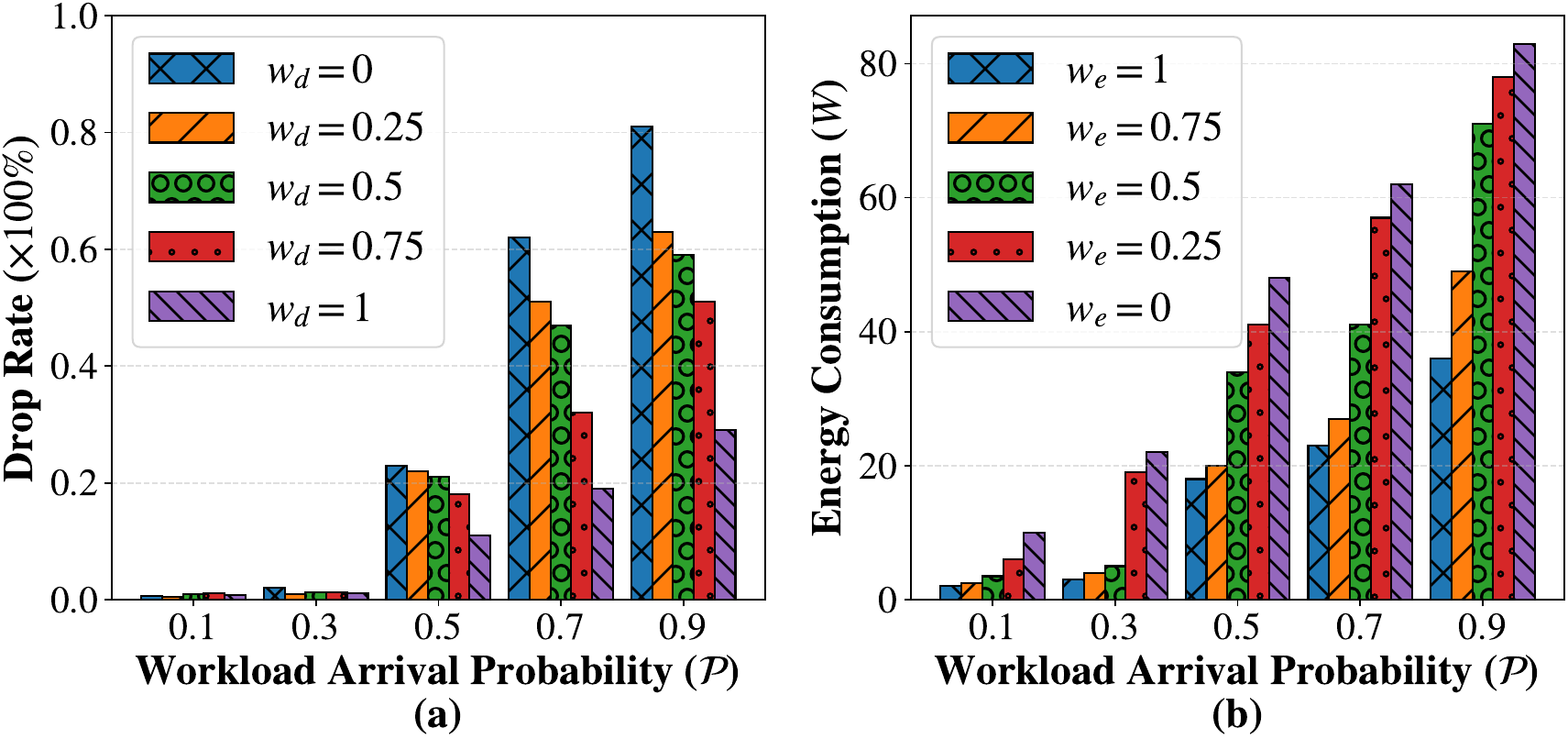}
\caption{DECOFFEE performance in terms of \textbf{(a)} drop rate and \textbf{(b)} energy consumption for different workload arrival probability ($\mathcal{P}$) and delay/energy awareness weights ($w_d$/$w_e$).}
\label{fig:res2}
\end{figure}

We first investigate the influence of the workload arrival probability $\mathcal{P}$, which reflects the traffic intensity generated by the IoT layer. Fig.~\ref{fig:res2} illustrates the DECOFFEE performance in terms of (a) percentage drop rate and (b) total energy consumption for different workload arrival probabilities and cost-awareness configurations. As expected, increasing $\mathcal{P}$ leads to higher system load, which consequently increases the probability of workload drops across all configurations (Fig.~\ref{fig:res2}a). When the delay-awareness coefficient $w_d$ is low (e.g., $w_d=0$ or $0.25$), the algorithm prioritizes energy efficiency and tends to avoid aggressive offloading decisions, which results in higher drop rates under heavy traffic. Conversely, higher values of $w_d$ lead the agents to prioritize latency-aware placement decisions, significantly reducing the drop rate even when the arrival probability approaches $\mathcal{P}=0.9$. In this sense, parameter $w_d$ can dynamically adapt workload placement policies depending on the selected delay–energy trade-off.

Fig.~~\ref{fig:res2}b presents the corresponding energy consumption behavior for the same workload arrival probabilities. In contrast to the drop rate results, energy consumption increases as the energy-awareness coefficient $w_e$ decreases. When the system prioritizes delay minimization (i.e., larger $w_d$ and smaller $w_e$), agents tend to perform more aggressive offloading or faster task processing, which results in higher overall energy consumption. Conversely, when the optimization focuses on energy efficiency (i.e., $w_e$ close to $1$), the agents prefer energy-conservative actions, resulting in significantly reduced power consumption. 

\subsubsection{Scalability to the Number of Agents} 

\begin{figure}[t]
\centering
\includegraphics[trim={0cm 0cm 0cm 0cm}, clip,width=1\columnwidth]{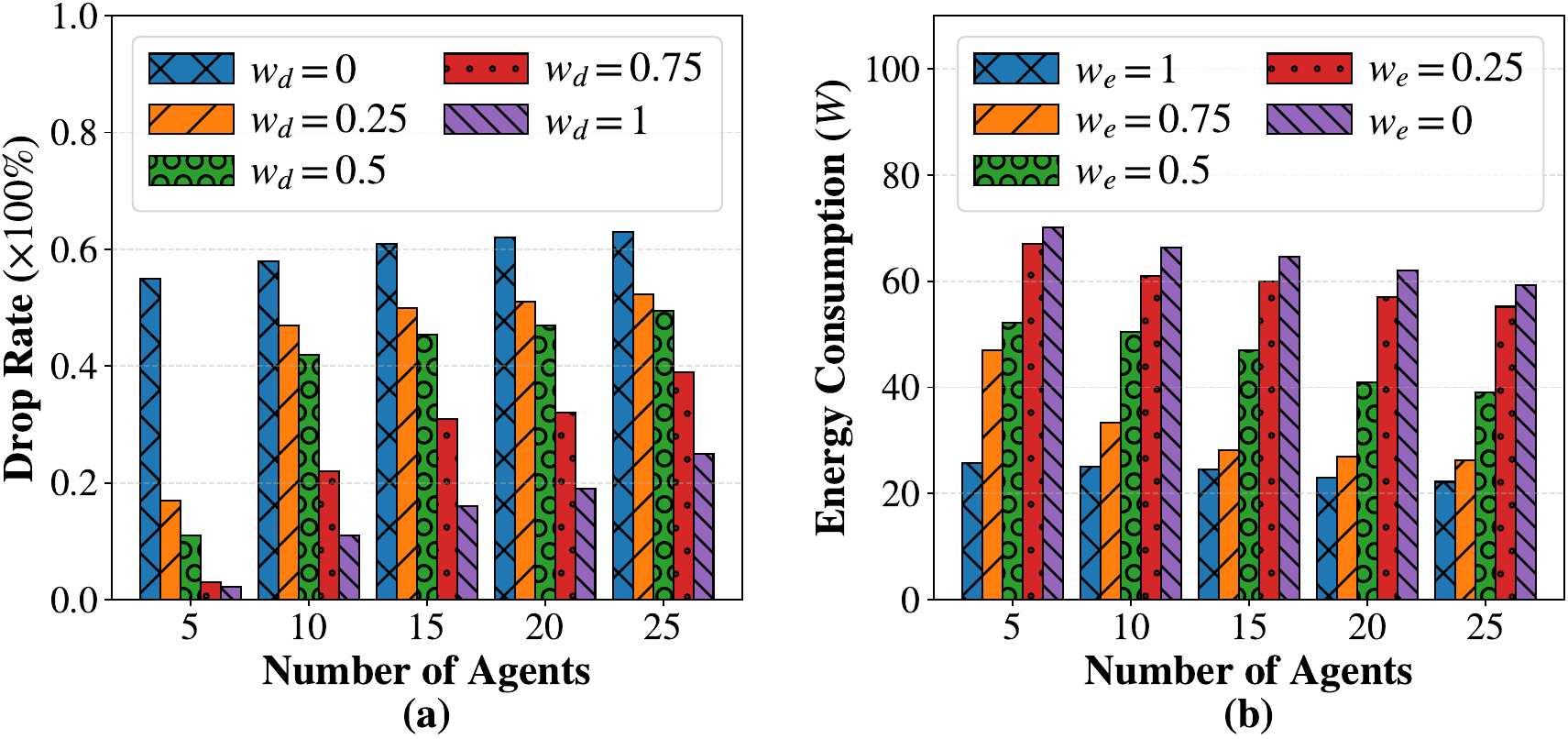}
\caption{DECOFFEE performance in terms of \textbf{(a)} drop rate and \textbf{(b)} energy consumption for increasing number of agents ($N$) and delay/energy awareness weights ($w_d$/$w_e$). Arrival probability is $\mathcal{P}=0.7$.}
\label{fig:res3}
\end{figure}

Next, the scalability of DECOFFEE as the number of distributed agents $N$ increases is analyzed. Fig.~\ref{fig:res3} illustrates the resulting drop rate and energy consumption for increasing network size under the same delay–energy awareness configurations, assuming a moderate heavy traffic intensity ($\mathcal{P}=0.7$).

From Fig.~\ref{fig:res3}(a), it can be observed that the drop rate generally increases with the number of agents when the delay-awareness coefficient is low. This occurs because a larger number of agents produces significant traffic increase and expands the state–action space, making coordination more complex and increasing the likelihood of suboptimal placement decisions under heavy traffic conditions. However, when delay awareness dominates (i.e., larger $w_d$ values), the algorithm maintains significantly lower drop rates even as the system scales, demonstrating reliable workload completion performance in larger distributed environments.

Fig.~~\ref{fig:res3}(b) illustrates the corresponding energy consumption as the number of agents increases. In contrast to the drop rate behavior, energy consumption tends to decrease slightly as the number of agents grows. This occurs because a denser Edge layer provides more offloading opportunities, allowing agents to distribute workloads more efficiently across available computing resources. However, configurations that prioritize delay minimization (small $w_e$ values) consistently exhibit higher energy consumption due to more aggressive workload processing decisions.

\subsubsection{Sensitivity to Computational Capacity}

\begin{figure}[t]
\centering
\includegraphics[trim={0cm 0cm 0cm 0cm}, clip,width=1\columnwidth]{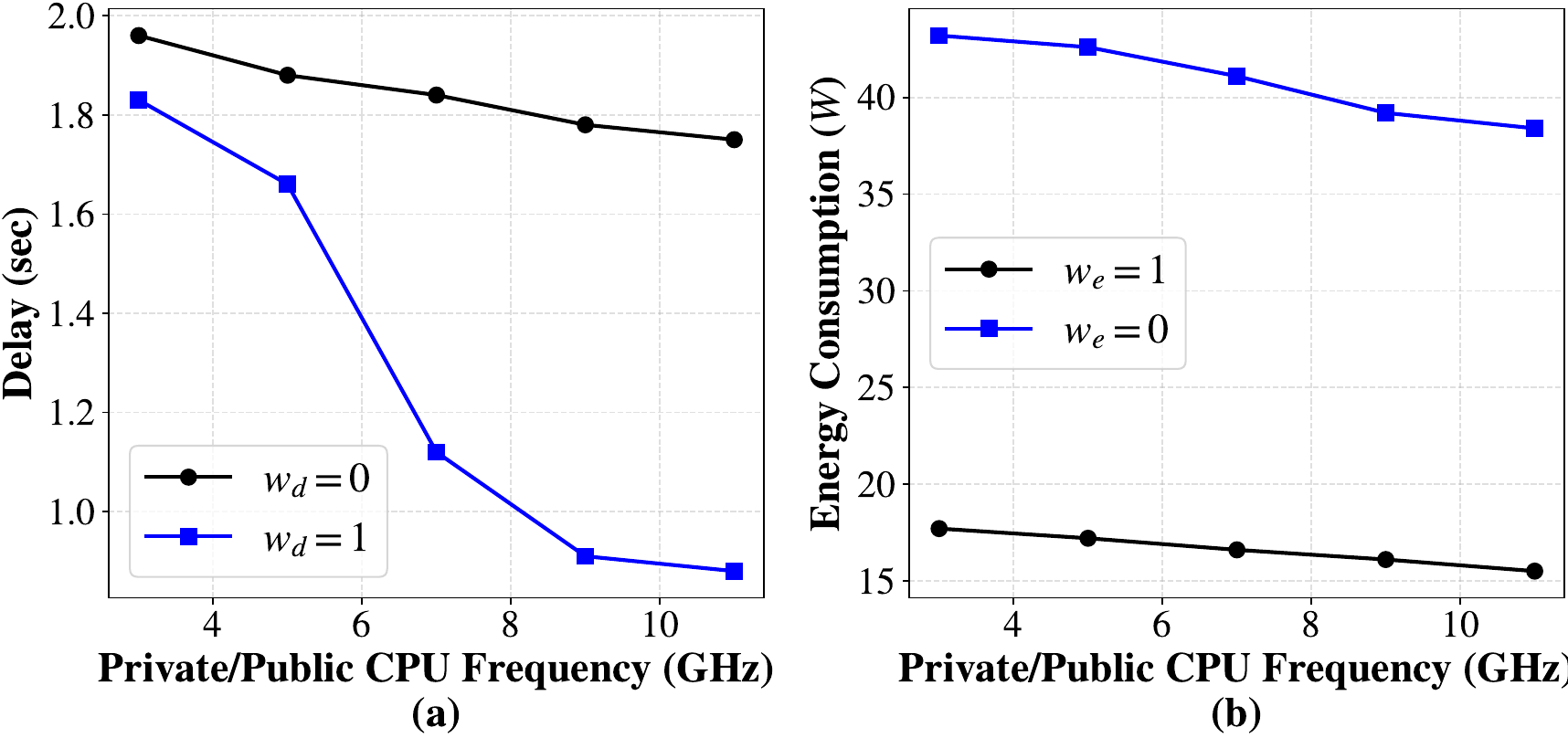}
\caption{DECOFFEE performance in terms of \textbf{(a)} average delay (sec) of executed workloads and \textbf{(b)} energy consumption for increasing capacity of private/public CPU (GHz) and delay/energy awareness.}
\label{fig:res4}
\end{figure}

We further investigate the influence of the available computational capacity of the edge computing nodes on the performance of DECOFFEE. Fig.~\ref{fig:res4} presents the average execution delay and the energy consumption as a function of the CPU processing capacity allocated to private and public workloads. Specifically, the CPU frequency varies from $3$ to $11$ GHz, representing different levels of computational capability at the Edge infrastructure. Fig.~\ref{fig:res4}(a) illustrates the average delay experienced by executed workloads under two extreme delay-awareness configurations. When the optimization focuses primarily on energy efficiency ($w_d=0$), the delay decreases only marginally as the available CPU capacity increases. This is because agents prefer local execution to avoid energy costs imposed by transferring the workloads, at the expense of increased waiting delays in the private WSs. In contrast, when delay awareness dominates ($w_d=1$), the average execution delay decreases significantly as CPU capacity increases, with the agents preferring more frequently to offload towards edge/cloud. This behavior indicates that the DRL agents actively exploit the additional processing resources to accelerate workload execution and reduce waiting delays when latency becomes a dominant optimization objective.

The corresponding energy consumption behavior is illustrated in Fig.~\ref{fig:res4}(b). In general, increasing CPU capacity leads to a gradual reduction in energy consumption due to shorter workload execution times and reduced waiting overhead. However, the absolute energy levels strongly depend on the energy-awareness coefficient. When energy minimization dominates ($w_e=1$), the overall power consumption remains significantly lower compared to the delay-dominant configuration ($w_e=0$), where the agents prioritize faster execution and offloading decisions at the expense of increased energy usage. In this context, it is evident that computational resources are exploited differently depending on the delay–energy trade-off.

\subsubsection{Distribution of Placement Decisions}

\begin{figure}[t]
\centering
\includegraphics[trim={0cm 0cm 0cm 0cm}, clip,width=1\columnwidth]{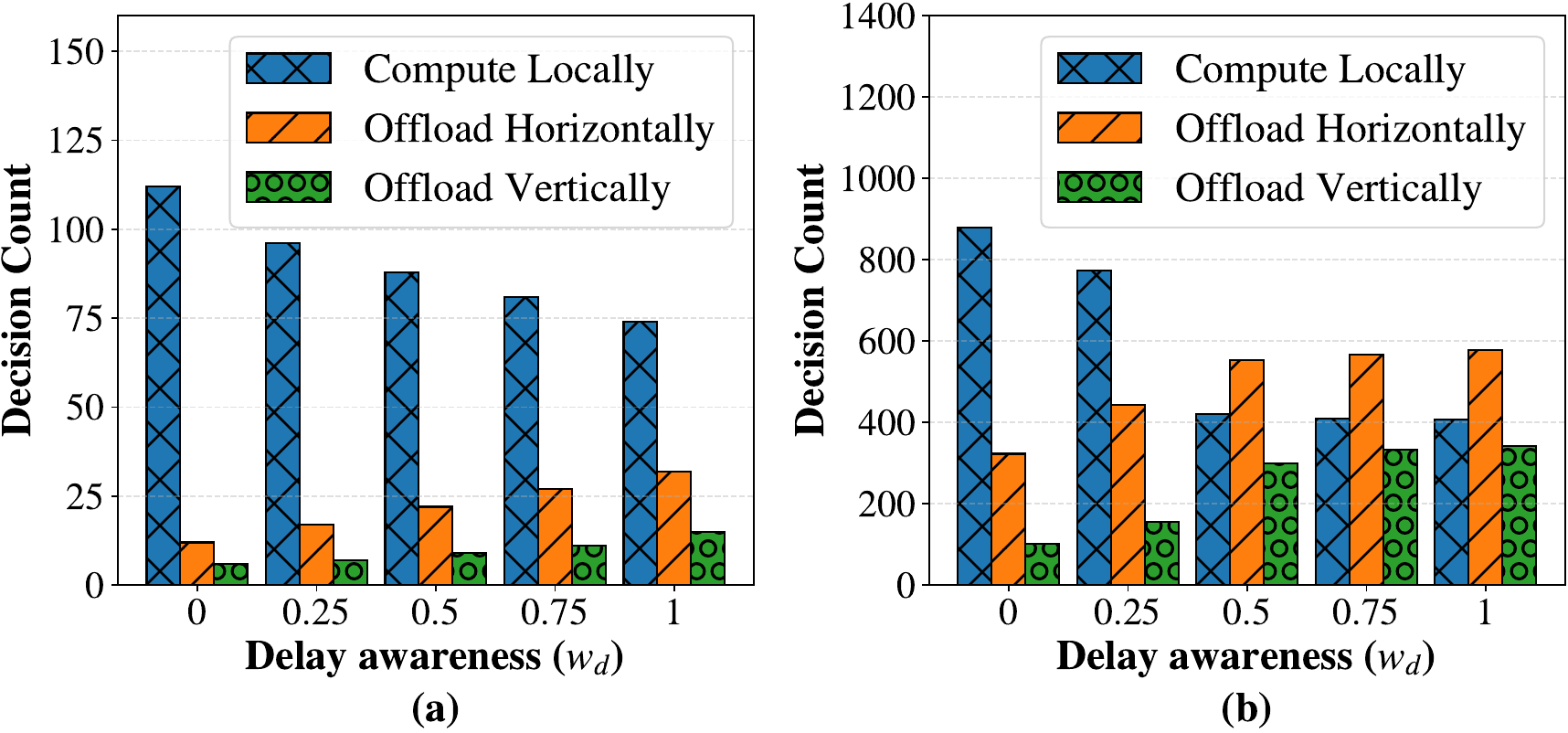}
\caption{Decision selection distribution across different delay/energy awareness weights under \textbf{(a)} sparse workload traffic ($\mathcal{P}=0.1$) and \textbf{(b)} dense traffic ($\mathcal{P}=0.7$).}
\label{fig:res5}
\end{figure}

To gain deeper insight into the behavior of the learned policies, Fig.~\ref{fig:res5} illustrates the distribution of decision selections across the three available workload placement options, namely local computation, horizontal and vertical offloading. The results are presented for two representative traffic conditions: sparse ($P=0.1$) and dense workload traffic ($P=0.7$). Under sparse workload traffic (see Fig.~\ref{fig:res5}(a)), the majority of tasks are processed locally when delay awareness is low. As the delay-awareness coefficient $w_d$ increases, the agents gradually shift toward horizontal offloading, distributing workloads among neighboring agents in order to reduce execution latency. Vertical offloading remains relatively limited in this scenario because the Edge infrastructure typically has sufficient resources to handle the incoming tasks.

Under dense workload traffic conditions (see Fig.~~\ref{fig:res5}(b)), the behavior changes significantly. For high energy awareness, DECOFFEE agents do not prefer frequent offloading decisions and local computation dominates. As the delay-awareness coefficient increases, both horizontal and vertical offloading decisions become more frequent. In particular, vertical offloading toward the Cloud becomes more prominent when delay awareness dominates, indicating that the agents increasingly exploit the larger computational capacity of the Cloud to prevent congestion in the Edge layer. At the same time, local computation decisions decrease because processing tasks locally under heavy load may lead to excessive waiting delays. As a result, the learned placement policies for different $w_d$ configurations can dynamically balance local processing and offloading decisions according to the current system load.

\subsection{Impact of LSTM inclusion}

\subsubsection{The Impact of LSTM Inclusion}

\begin{figure}[t]
\centering
\includegraphics[trim={0cm 0cm 0cm 0cm}, clip,width=1\columnwidth]{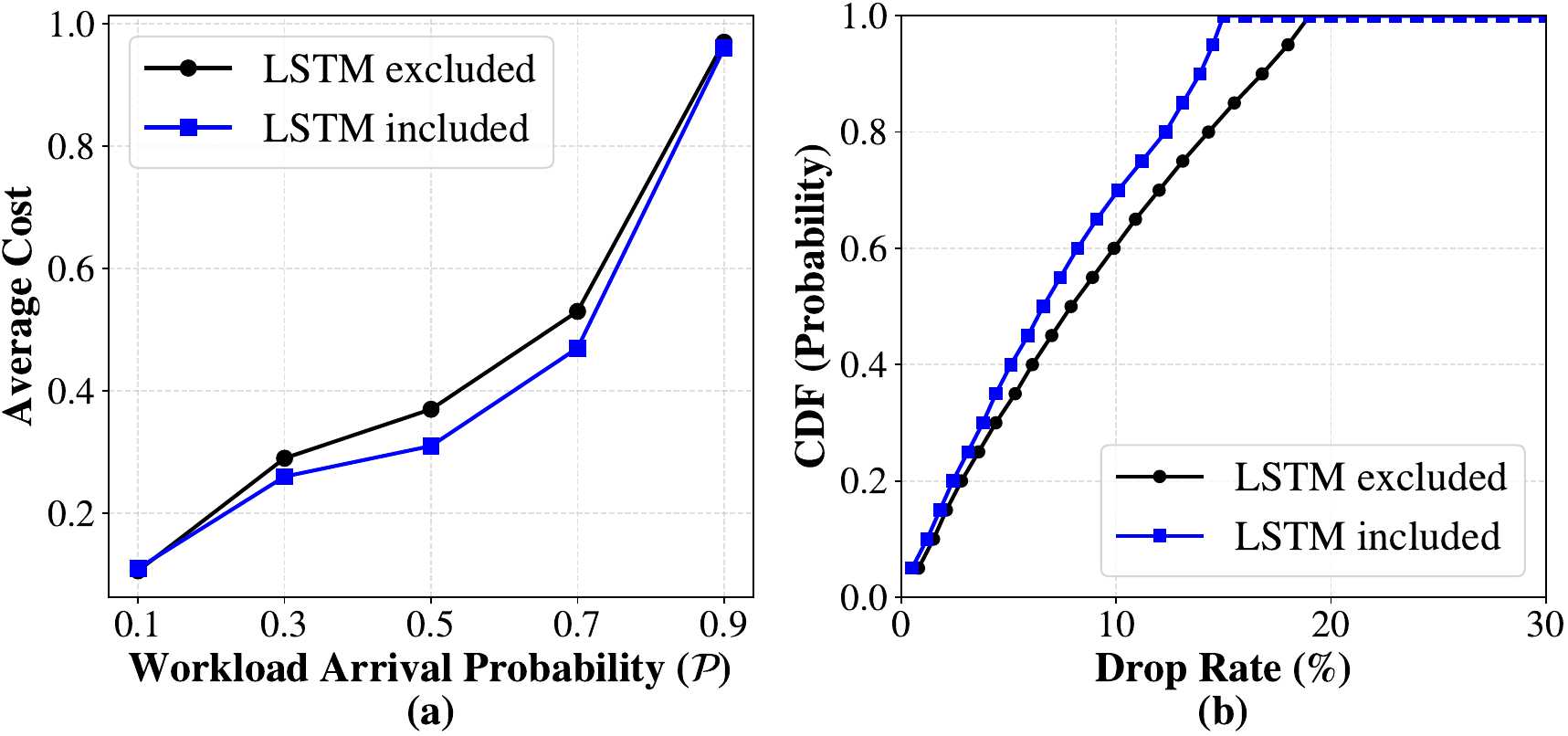}
\caption{DECOFFEE performance with \textit{vs} without LSTM. \textbf{(a)} Average cost (a.u.) for increasing workload probability function ($\mathcal{P}$). \textbf{(b)} CDF of the percentage Drop Rate. Delay-awareness weight is $w_d=1$.}
\label{fig:res6}
\end{figure}

The contribution of the LSTM-based forecasting module integrated into the DECOFFEE agents is also evaluated. The LSTM predictions are incorporated into the state representation of each DRL agent, allowing the offloading decisions to be proactive rather than purely reactive to the current system state. Fig.~\ref{fig:res6}(a) compares the average system cost achieved by DECOFFEE (after $100$ inference runs) with and without the LSTM prediction module for increasing workload arrival probability $P$ ($w_d=1$ for this comparison). It can be observed that the benefit of LSTM inclusion is most evident under intermediate traffic conditions, particularly for $\mathcal{P} \in [0.3,0.7]$. In this operating region, the LSTM-enabled DECOFFEE consistently achieves lower average cost than the version without forecasting, with an approximate reduction of about $10\%$ at $\mathcal{P}=0.3$, around $16\%$ at $\mathcal{P}=0.5$, and about $11\%$ at $P=0.7$. This indicates that, under moderate-to-high workload pressure, anticipating future congestion allows the agents to select more effective placement actions. In contrast, for very light traffic ($P=0.1$), both schemes perform similarly because the system remains far from saturation and forecasting offers limited additional benefit. Likewise, for extremely dense traffic ($P=0.9$), the two curves nearly converge, suggesting that the system approaches a saturated regime where even predictive placement has limited room for improvement.

Fig.~\ref{fig:res6}(b) further illustrates the cumulative distribution function (CDF) of the percentage drop rate for the two configurations, considering $w_d=1$ and $\mathcal{P}=0.5$. The curve corresponding to the LSTM-enabled DECOFFEE shifts toward lower drop-rate values, indicating improved reliability in workload execution. In particular, the maximum drop rate is around 15\% for LSTM-enabled DECOFFEE, whereas that of LSTM-free approaches 19\%. This confirms that incorporating temporal prediction into the DRL state representation improves the robustness of the learned policy.

\subsection{Baseline Comparisons}

To thoroughly assess the performance of the DECOFFEE framework, we compare it against several representative baseline workload placement strategies under different configurations. The considered baseline schemes reflect both simple rule-based policies and more advanced delay-oriented approaches commonly used in the task offloading literature \cite{giannopoulos2024hoodie, tang2020deep, liu2016delay}. The evaluated schemes are summarized as follows:

\begin{enumerate}
    \item \textbf{Random Placement (RP)}: Each agent randomly selects one of the available actions with equal probability among local computation, vertical offloading to the Cloud, and horizontal offloading to another EA. In the case of horizontal offloading, the destination EA is selected uniformly at random with probability $1/(N-1)$.

    \item \textbf{Local-Only Processing (LOP)}: Each agent executes all incoming workloads locally without performing any form of offloading. This scheme represents the most conservative placement strategy and serves as a lower-bound baseline when distributed resources are not utilized.

    \item \textbf{Cloud-Only Offloading (COO)}: Each agent forwards all incoming workloads to the Cloud for processing. This approach reflects scenarios where centralized computing resources dominate the workload processing, but may introduce additional communication latency and network congestion.

    \item \textbf{Edge-to-Edge Offloading (EEO)}: Each agent offloads all workloads horizontally to neighboring EAs. The destination EA is selected randomly with probability $1/(N-1)$. This scheme represents an extreme case where the Edge layer handles all workloads collaboratively without utilizing Cloud resources.

    \item \textbf{Round-Robin Offloader (RRO)}~\cite{alhaidari2021enhanced}: Each agent distributes incoming workloads across the available placement options in a cyclic manner. For example, considering EA $n=1$, the first workload is executed locally, the second is offloaded to the Cloud, the third is sent to EA $2$, the fourth to EA $3$, and so forth. This strategy ensures balanced utilization of computing nodes but does not adapt to dynamic system conditions.

    \item \textbf{Minimum Latency Estimation Offloader (MLEO)}~\cite{liu2018offloading, liu2016delay}: In this powerful scheme, each agent estimates the expected execution delay for every available placement option and selects the option that minimizes the predicted latency. Specifically, upon arrival of a new workload $w_n(t)$ at time slot $t$, agent $n$ computes $N+1$ delay estimates corresponding to the available destinations. The local execution delay is obtained directly using the private WS delay model, while the delay associated with offloading to another node $k \in \mathcal{N}-\{n\}$ is calculated by combining the offloading delay and the estimated waiting time in the public queue of node $k$. The agent then selects the placement option that yields the lowest estimated delay.

    \item \textbf{DECOFFEE}: Three variants of DECOFFEE scheme are considered for comparisons, including the delay-aware (DA-DECOFFEE) with $w_d=1$, the energy-aware (EA-DECOFFEE) with $w_e=1$, and the balanced (B-DECOFFEE) with $w_d=w_e=0.5$. 
\end{enumerate}

For all schemes, the performance is evaluated using three key metrics: the average workload execution delay (of the successfully processed workloads), the percentage drop ratio, and the overall energy consumption. The reported results are obtained by averaging the performance across $200$ validation episodes after training convergence.

To analyze the delay performance, a relatively relaxed workload deadline of $10$ seconds is adopted, allowing tasks to remain in the system long enough to capture the execution latency of successfully processed workloads. In contrast, when evaluating the drop ratio, a stricter deadline of $2$ seconds is considered, making deadline violations more likely and therefore highlighting the robustness of the workload placement strategies under tight time constraints.

\begin{figure}[t]
\centering
\includegraphics[trim={0cm 0cm 0cm 0cm}, clip,width=1\columnwidth]{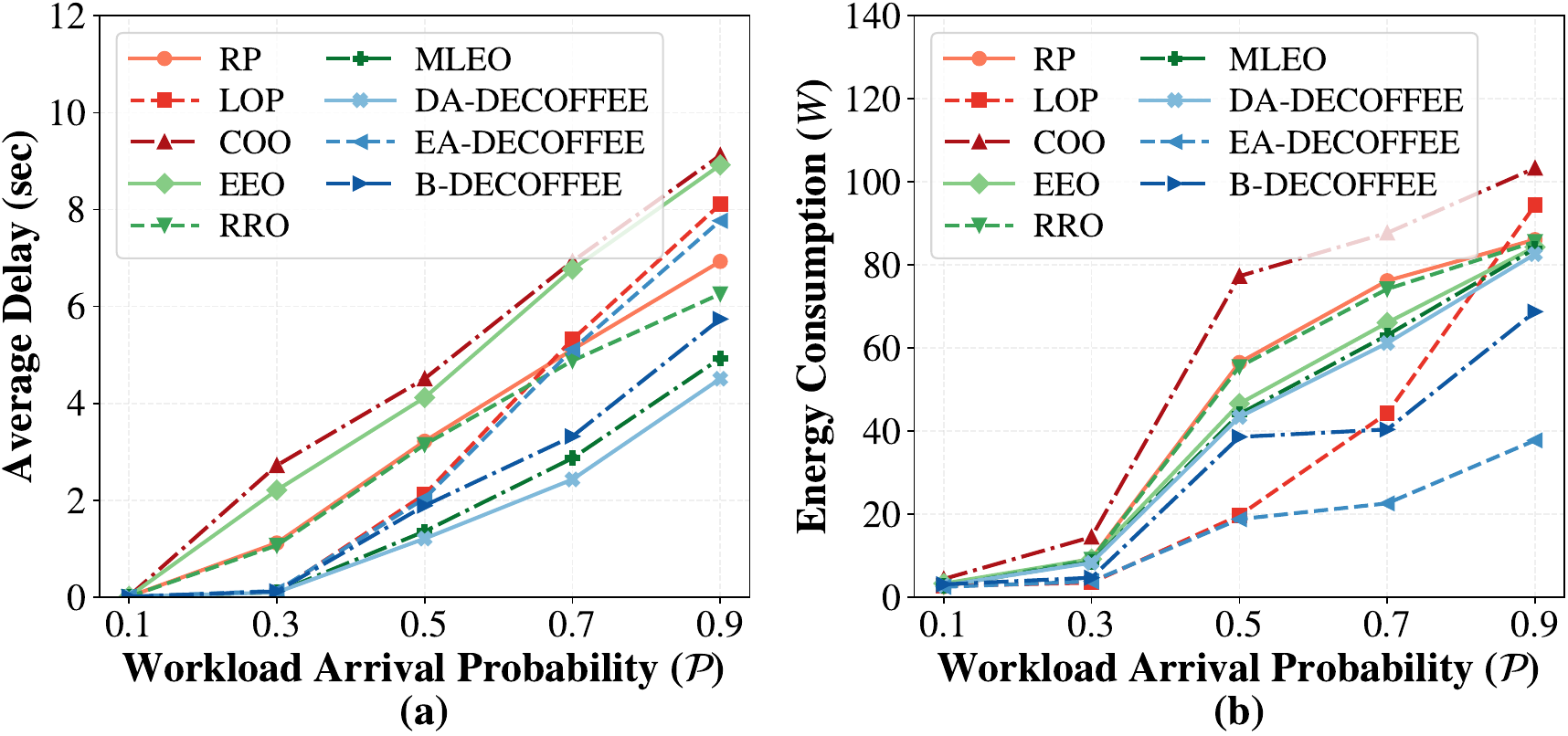}
\caption{Performance comparison among the 9 workload placement schemes in terms of \textbf{(a)} Average delay (sec) and \textbf{(b)} energy consumption (W) for increasing arrival probability ($\mathcal{P}$).}
\label{fig:res7}
\end{figure}

\begin{figure}[t]
\centering
\includegraphics[trim={0cm 0cm 0cm 0cm}, clip,width=1\columnwidth]{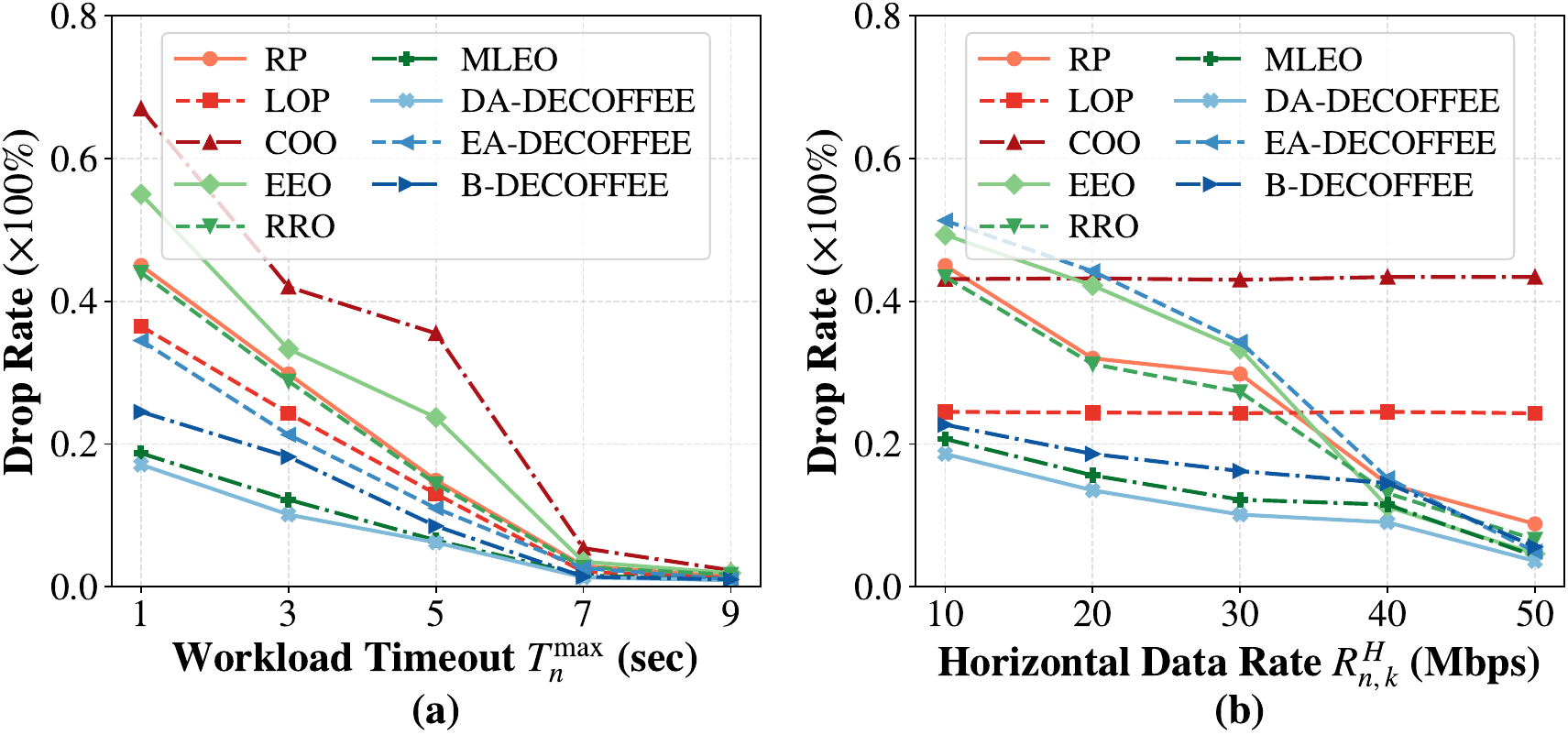}
\caption{Drop rate (\%) achieved by the 9 workload placement schemes for \textbf{(a)} increasing workload timeout values ($T_n^{\max}$, in sec) and \textbf{(b)} increasing data-rate of horizontal links ($R_{n,k}^H$, in Mbps). Arrival probability is $\mathcal{P}=0.5$.}
\label{fig:res8}
\end{figure}

Fig.~\ref{fig:res7} illustrates the comparative performance of all workload placement schemes for increasing workload arrival probability. Specifically, Fig.~\ref{fig:res7}(a) presents the average execution delay, while Fig.~\ref{fig:res7}(b) reports the total energy consumption. As expected, the average delay increases with the workload arrival probability across all schemes due to the higher congestion levels in the system. More specifically, the delay-aware variant (DA-DECOFFEE) achieves the lowest delay across all traffic intensities, demonstrating the effectiveness of the learned placement policy in minimizing waiting and execution delays. For instance, at high workload intensity ($\mathcal{P}=0.9$), DA-DECOFFEE reduces the average delay by approximately $8\%$ compared to the minimum-latency heuristic (MLEO) and by more than $50\%$ compared to the cloud-only offloading strategy (COO). The balanced variant (B-DECOFFEE) also maintains significantly lower delay than most baselines while achieving a more favorable delay–energy trade-off. In contrast, rule-based strategies such as RP, LOP, and RRO exhibit significantly higher delays as the workload intensity increases, since these schemes lack the ability to dynamically adapt workload placement decisions according to the system state.

Fig.~\ref{fig:res7}(b) further compares the schemes in terms of energy consumption. The energy-aware DECOFFEE variant (EA-DECOFFEE) consistently achieves the lowest power consumption across all workload intensities. For example, at $\mathcal{P}=0.9$, EA-DECOFFEE reduces the total energy consumption by roughly $60\%$ compared to the cloud-only offloading scheme and by about $45\%$ compared to the minimum-latency heuristic. Notably, B-DECOFFEE variant exhibits the third best and second best solution in terms of delay and energy, respectively, guaranteeing a beneficial delay-energy trade-off across all schemes.

To further evaluate the robustness of the compared schemes, Fig.~\ref{fig:res8} examines the drop rate performance under different system parameters when $\mathcal{P}=0.5$. Fig.~\ref{fig:res8}(a) shows the drop rate as a function of the workload timeout $T_n^{\max}$, while Fig.~\ref{fig:res8}(b) evaluates the drop rate for increasing horizontal offloading data rate $R^H_{n,k}$. From Fig.~Fig.~\ref{fig:res8}(a), it is evident that increasing the workload timeout significantly reduces the percentage of dropped workloads across all schemes, as tasks are given more time to complete execution. Nevertheless, the DA-DECOFFEE variant consistently achieve the lowest drop rates across the entire range of timeout values, with MLEO yielding slightly higher drop rates. Noteworthy, as the $T_n^{\max}$ becomes more stringent, the performance separation across schemes becomes more obvious. Furthermore, the B-DECOFFEE demonstrates the third best solver across all timeout values, following DA-DECOFFEE and MLEO schemes. In contrast, schemes that rely on rigid placement policies, such as COO and LOP, exhibit significantly higher drop ratios, especially under strict timeout constraints.

Fig.~\ref{fig:res8}(b) evaluates the effect of the horizontal communication capacity on workload execution reliability. As the horizontal data rate increases, the drop rate decreases for most schemes because tasks can be transferred more efficiently among Edge agents. Again, the DA-DECOFFEE variant outperforms all baseline schemes across the entire range of communication capacities. MLEO and B-DECOFFEE are again the second and thrid best solutions for all the tested horizontal rates. Notably, the EEO scheme shows a significant decrease in achieved drop rates as the horizontal data rate increase because delays imposed by horizontal transfers are significantly low. The results confirm that DA-DECOFFEE can be the optimal selection when workload placement latency is of importance, whereas EA-DECOFFEE can be used for energy savings when workload timeout is more flexible. By leveraging distributed DRL with predictive system awareness, DECOFFEE can dynamically adapt workload placement decisions to changing system conditions, achieving lower delay, reduced energy consumption, and improved task completion reliability.

%
%
%
%
\section{Conclusion and Future Extensions}\label{sec:conclusions}

\subsection{Conclusion}

This paper presented DECOFFEE, a decentralized reinforcement learning framework for delay- and energy-aware workload placement across the Edge–Cloud computing continuum. The proposed approach models the workload offloading process as parallel Markov Decision Processes and employs a Double Dueling Deep Q-Network architecture enhanced with LSTM-based forecasting to proactively capture future system conditions. By enabling each edge node to operate as an autonomous learning agent, DECOFFEE dynamically adapts workload placement decisions according to the observed system state and predicted resource utilization. 

Extensive numerical evaluations demonstrated that DECOFFEE significantly outperforms conventional rule-based and heuristic workload placement schemes in terms of execution delay, energy consumption, and workload drop rate. The results further showed that the framework remains robust across varying workload intensities, network capacities, and system scales, while allowing flexible trade-offs between latency and energy efficiency through configurable cost weights. Overall, the proposed decentralized learning architecture provides a scalable and adaptive solution for intelligent workload management in future Edge–Cloud infrastructures.

\subsection{Future Directions}

Several research directions emerge from this work. First, future studies may directly investigate federated or collaborative learning mechanisms to allow distributed agents to share knowledge while preserving system scalability and privacy. Second, extending the framework to support heterogeneous application classes with different quality-of-service requirements and resource profiles would further enhance its applicability to real-world IoT ecosystems. Third, incorporating additional system dimensions such as network congestion, communication reliability, and carbon-aware computing policies could improve the sustainability and operational efficiency of workload placement decisions. Finally, validating the DECOFFEE framework on real-world Edge–Cloud testbeds and integrating it with orchestration platforms for practical deployment remains an important step toward operational adoption.

\section*{Acknowledgments}
The authors would like to thank I. Paralikas for his contribution.

\bibliographystyle{IEEEtran}

\bibliography{bibliography}

 
\vspace{11pt}

\begin{IEEEbiography}[{\includegraphics[width=1in,height=1.25in]{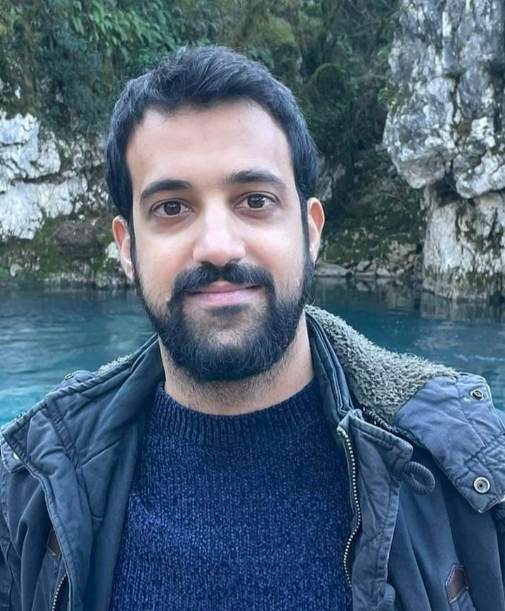}}]{ANASTASIOS E. GIANNOPOULOS}~(Member, IEEE) (M.Eng, Ph.D) received the diploma of Electrical and Computer Engineering from the National Technical University of Athens (NTUA), where he also completed his Master Engineering (M.Eng) degree, in 2018. He also obtained his Ph.D. at the Wireless and Long Distance Communications Laboratory of NTUA. His research interests include advanced Optimization Techniques for Wireless Systems, ML-assisted Resource Allocation, Maritime Communications and Multi-dimensional Data Analysis.

He is currently working as a Research Associate at the National and Kapodistrian University of Athens, as well as Post-Doc at National Technical University of Athens. He has authored more than 75 scientific publications in the fields of Wireless Network Optimization, Computing Continuum, Machine Learning and Brain Multi-dimensional Analysis. Since 2022, he is a Member of IEEE and reviewer in several IEEE journals (IEEE Transactions on Mobile Computing, IEEE Vehicular Technology Magazine, IEEE Network, IEEE Access).
\end{IEEEbiography}

\begin{IEEEbiography}[{\includegraphics[width=1in,height=1.25in]{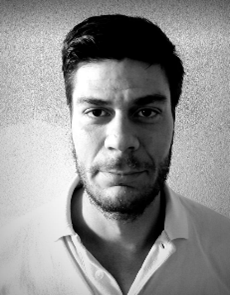}}]{SOTIRIOS T. SPANTIDEAS}~(D.Eng, M.Sc, Ph.D) obtained the Diploma of Electrical $\&$ Computer Engineering from the Polytechnic School of the University of Patras in 2010. He then attended the Master Program "Electrophysics" at the Royal Institute of Technology in Stockholm (KTH), from which he obtained the title MSc in 2013. In 2018 he obtained his PhD from the National Technical University of Athens (NTUA) with doctoral dissertation entitled ``Development of Methods for obtaining DC and low frequency AC magnetic cleanliness in space missions". His research interests include Electromagnetic Compatibility, Machine Learning for Wireless Networks, Magnetic Cleanliness for space missions and optimization algorithms for Computational Electromagnetics.

From 2014, he is working as a Research Associate with NTUA and the National and Kapodistrian University of Athens (Department of Ports Management and Shipping - NKUA), participating in multiple Horizon projects. He has published over 50 papers in scientific journals and conferences in the fields of Electromagnetic Compatibility, Optimization Methods for Wireless Networks and Machine Learning for Resource Allocation problems.
\end{IEEEbiography}

\begin{IEEEbiography}[{\includegraphics[width=1in,height=1.25in]{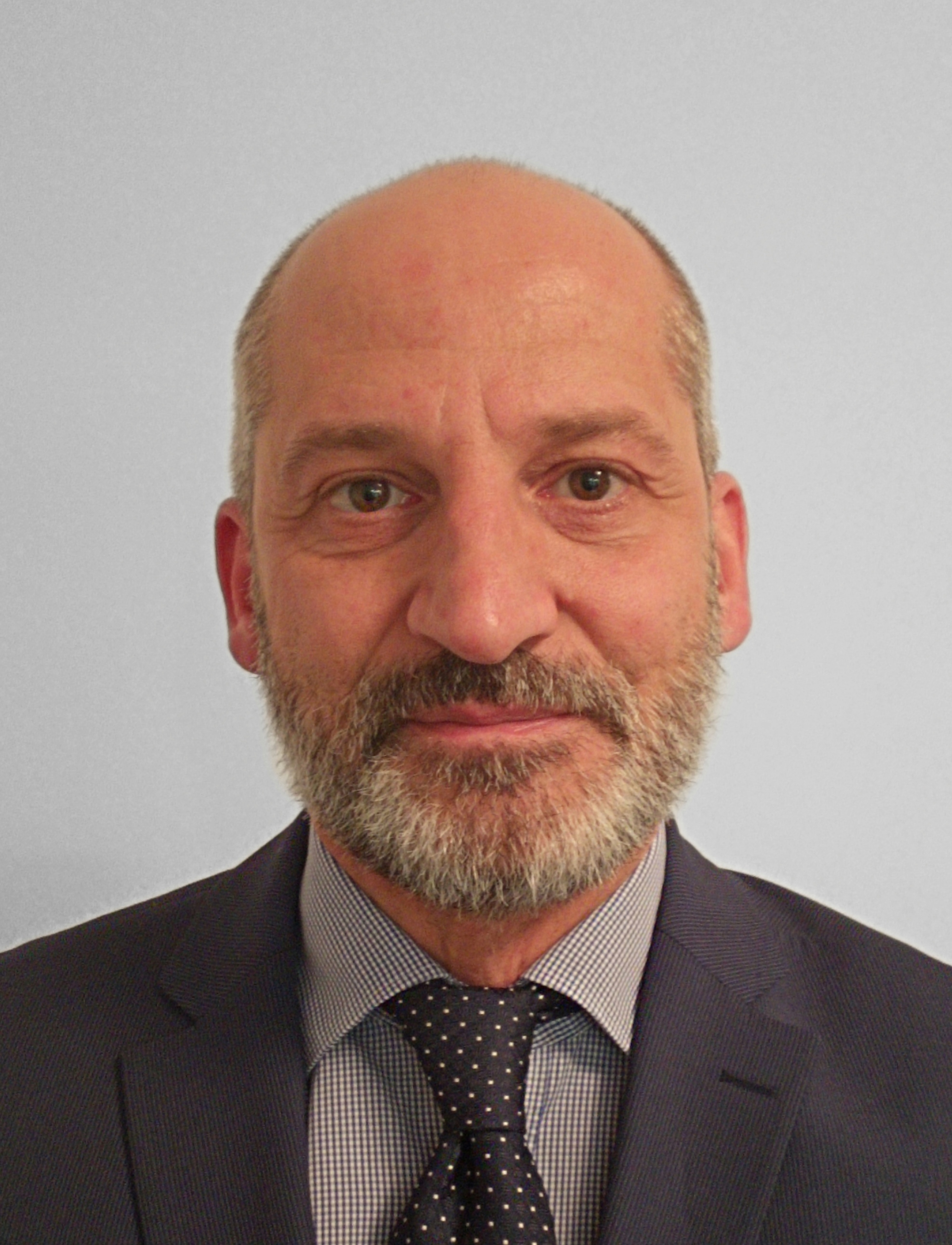}}]{PANAGIOTIS TRAKADAS}~(MEng, Ph.D) received his Diploma, Master Engineering, Degree in Electrical $\&$ Computer Engineering and his Ph.D. from the National Technical University of Athens. His main interests include 5G/6G technologies, such as O-RAN, NFV and NOMA, Wireless Sensor Networks, and Machine Learning based Optimization Techniques for Wireless Systems and Maritime Communications. 

He is currently an Associate Professor at the Department of Ports Management and Shipping, in National and Kapodistrian University of Athens and has been actively involved in many national and EU-funded research projects. He has published more than 180 papers in magazines, journals, books and conferences and a reviewer in several journals and TPC in conferences.
\end{IEEEbiography}

\vfill

\end{document}